\documentclass[12pt]{iopart}
\usepackage{iopams}

\usepackage[unicode=true,pdfusetitle, bookmarks=true,bookmarksnumbered=false,
bookmarksopen=true, breaklinks=false,pdfborder={0 0 0},backref=false,
colorlinks=true, linkcolor=myrefcolor,citecolor=myurlcolor,urlcolor=myurlcolor]{hyperref}

\usepackage{graphicx}
\usepackage{cite}
\usepackage{color}
\usepackage{bbm}
\usepackage{dsfont}
\newcommand{\upe}{\mathrm{e}}
\newcommand{\upi}{\mathrm{i}}

\newcommand{\Moplus}{\bigoplus}
\renewcommand{\tr}{\mathrm{Tr}}
\newcommand{\ket}[1]{{\left|{#1}\right\rangle}}
\newcommand{\bra}[1]{{\left\langle{#1}\right|}}

\newcommand{\binom}[2]{{\left(\begin{array}{@{}c@{}}#1\\#2\end{array}\right)}}
\newcommand{\id}{\mathds{1}}
\newcommand{\Ntot}{N_{\mathrm{total}}}
\newcommand{\rhoN}{\rho_N}
\newcommand{\thetahatN}{\hat{\theta}_N}
\newcommand{\FNtheta}{F_{N;\theta}}
\newcommand{\FQNtheta}{\mathcal{F}_{N;\theta}}
\newcommand{\FQNeps}{\mathcal{F}_{N;\varepsilon}}
\newcommand{\FQNphi}{\mathcal{F}_{N;\varphi}}
\newcommand{\thetavec}{\boldsymbol{\theta}}
\newcommand{\FQNthetavec}{\mathcal{F}_{N;\thetavec}}
\newcommand{\thetahatNvec}{\hat{\thetavec}_N}
\newcommand{\cmatN}{\mathcal{C}_N}
\renewcommand{\Re}{\mathrm{Re}}
\renewcommand{\Im}{\mathrm{Im}}
\newcommand{\Jvec}{\boldsymbol{J}}
\newcommand{\cP}{\mathcal{P}}
\newcommand{\expt}[1]{{\left\langle{#1}\right\rangle}}

\renewcommand{\th}{\textsc{h}}
\newcommand{\tv}{\textsc{v}}
\newcommand{\vac}{\textsc{vac}}

\newcommand{\proj}[1]{\ket{#1}\!\bra{#1}}
\newcommand{\ad}[2]{a^\dagger_{\footnotesize{#1},#2}}
\renewcommand{\aa}[2]{a_{\footnotesize{#1},#2}}
\newcommand{\bd}[2]{b^\dagger_{\footnotesize{#1},#2}}

\newcommand{\sig}[1]{\mathsf{S}_{#1}}
\newcommand{\pr}[1]{\mathsf{Pr}({#1})}

\definecolor{myurlcolor}{rgb}{0,0,0.7}
\definecolor{myrefcolor}{rgb}{0.8,0,0}

\begin{document}
	
	\title[ ]{Multiparameter estimation for qubit states with collective measurements: a case study}
	
	\author{Yink Loong Len}
	
	\address{Centre for Quantum Optical Technologies, Centre of New Technologies, University of Warsaw, \\ Banacha 2c, 02-097, Warsaw, Poland\\
	y.len@cent.uw.edu.pl}
	\vspace{10pt}
	\begin{indented}
		\item[] \today
	\end{indented}
	
	\begin{abstract}
		Quantum estimation involving multiple parameters remains an important problem of both theoretical and practical interest. In this work, we study the problem of simultaneous estimation of two parameters that are respectively associate with the length and direction of the Bloch vector for identically prepared qubit states that is confined to a plane, where in order to obtain the optimal estimation precision for both parameters, collective measurements on multiple qubits are necessary. Upon treating $N$ qubits as an ensemble of spin-1/2 systems, we show that simultaneous optimal estimation for both parameters can be attained asymptotically with a simple collective measurement strategy---First, we estimate the length parameter by measuring the populations in spaces corresponding to different total angular momentum values $j$, then we estimate the direction parameter by performing a spin projection onto an optimal basis. Furthermore, we show that when the state is nearly pure, for sufficiently but not arbitrarily large $N$, most information will be captured in the largest three $j$-subspaces. Then, we study how the total angular-momentum measurement can be realized by observing output signatures from a Bell multiport setup, either exactly for $N=2,3$, or approximately when the qubits are nearly pure for other $N$ values. We also obtain numerical results that suggest that using a Bell multiport setup, one can distinguish between projection onto the $j=N/2$ and $j=N/2-1$ subspaces from their respective interference signatures at the output.
	\end{abstract}

\section{Introduction}	
An important aspect in physical and information sciences is the acquisition of optimal knowledge about some quantities of interest, through the acts of measurement on the information carriers or probes. Quantum parameter estimation and quantum metrology study, in particular, the ultimate attainable precision in estimating the parameters of interest, under the constraints set by quantum theory. Estimation of a relative phase, such as one in an optical interferometer for gravitational wave sensing \cite{Abbott2016,Abbott2017,Tse2019} or atomic states for frequency measurements and magnetic field sensing \cite{Bollinger1996,Zhang2016,Maze2008,Taylor2008}, temperature in quantum systems \cite{Kucsko2013,Mehboudi2019,Moreva2020}, as well as the spatial extent of composite light sources\cite{Ram2006,Donnert2006,Raj2008,Weissleder2008} in imaging tasks, are examples in which the ultimate precise estimators are highly desired. 

While it is quite typical to have just a single parameter of interest, more generally, one might encounter quantum estimation problems that concern genuinely multiple parameters, and for which \emph{simultaneous} or \emph{joint}-parameter estimation is called for \cite{Szczykulska2016,Liu2019,Demkowicz2020}. They include, for instance, estimation of phases and decoherence strength\cite{Vidrighin2014,Crowley2014}, numerous parameters for unitary operations\cite{Humphreys2013,Baumgratz2016,Goldberg2020}, and all or some parameters characterizing the spatial extent of light sources, such as the centroid, brightness, separation and orientation \cite{Tsang2016,Nair2016,Lupo2016}. Understandably, finding and implementing the optimal measurement for joint estimation of multiple parameters is generally more challenging than in the case of single-parameter estimation. In particular, joint estimation need not be ``\emph{compatible}",
i.e., unless certain conditions are met, using no matter what measurement and estimation strategy, one can never \emph{simultaneously} estimate \emph{all} the parameters 
as precisely as one would optimally achieve when estimating just one parameter at a time, while assuming the other parameters are known and fixed.
This is not surprising, as the measurement bases needed to attain the optimal precision for different parameters need not correspond to complementary observables\cite{Bohr1928,Schwinger1960}. 

Many important aspects and interesting results on multi-parameter estimation have been established. They include: the general conditions for ``compatible" joint estimation\cite{Ragy2016}, the mathematical conditions for the measurement operators being optimal in estimating all the parameters \cite{Pezze2017,Yang2019}, and bounds on estimation precision and their evaluations \cite{Holevo1982,Yuen1973,Nagaoka1989,Gill2000,Hayashi2005,Guta2006,Hayashi2008,Kahn2009,Yamagata2013,Suzuki2016, Carollo2019, Albarelli2019,Tsang2020, Sidhu2021,Yamagata2021,Albarelli2021,Conlon2021}. However, given a specific multi-parameter quantum estimation problem, the measurement that attains the optimal joint estimation precision is still not explicitly deducible and recognizable, not to say its implementation scheme. Finding them remains thus a problem of both theoretical and practical interest.

In this work, we perform a case study on joint estimation of parameters that respectively define the length and the direction of the Bloch vector for an ensemble of identically prepared qubit states. In particular, the Bloch vector is assumed to be confined to a known plane, such that we are here essentially studying a two-parameter estimation problem: one for specifying the length, another one for specifying the angle within the plane. Within the larger context for multiparameter estimation problem, here are looking at a specific model which is known as asymptotically classical \cite{Suzuki2019}, and moreover, that the two parameters are globally orthogonal, such that their joint estimation is compatible in the large ensemble limit. Despite its simplicity and being specific, however, this elementary model has significance for various physical problems, such as in quantum sensing and quantum imaging. For example, joint estimation of a phase shift and the amplitude of phase diffusion\cite{Vidrighin2014,Crowley2014}, states parameters in two-level quantum mechanical systems such as the purity and the phase for the polarization degrees of freedom of light \cite{James2001}, and the separation and centroid of two binary light sources which serves as a model to discuss resolution limits in imaging\cite{Chrostowski2017}, can be fittingly mapped into our model under appropriate circumstances. Hereinafter, we will not make explicit references to the physical meaning of the parameters, keeping the estimation problem at the abstract level. We also take notice of two related and detailed work on studying optimal full estimation of qubit states \cite{Hayashi2008,Bagan2006}, but with a somewhat different analysis and focus than what we will have here.

The organization of this work is as follows. In Sec.~\ref{Sec2}, we review the basic ingredients of quantum parameter estimation that we will use in this work. Readers that are familiar with the relevant concepts and results in quantum parameter estimation can then proceed directly to Secs.~\ref{Sec3} and \ref{Sec4}, where we formally introduce and study our two-parameter estimation problem. As a result, upon treating the qubits as a collection of spin-1/2 elementary systems, we show how the compatible joint estimation can be achieved in the large ensemble limit by the global or collective measurement of the total angular momentum squared operator, followed by a commuting, local projective measurement. Furthermore, we look deeper into the special case of nearly-pure qubit estimation, and highlight the link with the problem of sub-Rayleigh resolution imaging. We then discuss how such estimation scheme can be carried out in principle, either exactly or approximately, and by using either passive unitary setup or more general quantum circuits. From the obtained numerical results, we propose that the projection onto the two largest total angular momentum values could be achieved using a Bell multiport setup. Finally, we discuss briefly a few points in Sec.~\ref{Sec5} and conclude in Sec.~\ref{Sec6}.

\section{Preliminaries}\label{Sec2}
For the sake of completeness, in this section we review the basic ingredients of quantum parameter estimation theory that will be relevant for this work. We start with the single-parameter estimation scenario, then followed by multi-parameter estimation scenario.

\subsection{Single-parameter estimation}
\noindent  In this scenario, our task is to estimate the parameter, labeled as $\theta$, that is encoded in the qudit quantum state $\rho(\theta)$. Generally, we obtain information about, and hence an estimate for $\theta$, from the measurement statistics of $\Ntot$ independent and identical copies of $\rho(\theta)$. In particular, we consider $\Ntot=\nu N$, whereby we group $N$ copies together as an $N$-qudit ensemble described by the state $\rhoN(\theta)=\rho(\theta)^{\otimes N}$, and measure them with a probability-operator measurement (POM), specified by the set of $N$-qudit operators $\{M_\ell\}$, with $M_\ell\geq0\, \forall \ell,\; \sum_\ell M_\ell=\id_{d^N}$. From a single measurement, a single outcome $k$ will be recorded, with the probability of getting outcome $k=\ell$ given by the Born's rule, i.e., $p_\ell(\theta)=\Tr\{\rhoN(\theta) M_\ell\}$. Then, upon repeating over $\nu$ rounds of measurement, we obtain a sequence of $\nu$ measurement outcomes $\{k_1,k_2,\cdots,k_\nu\}$ as our data $D$, from which we construct an estimate\footnote{The estimator $\thetahatN$ depends of course on $\nu$ as well. Here we write out only the dependence on $N$ but not $\nu$ explicitly to avoid overloading the notation.}  $\thetahatN(D)$, using for example the maximum-likelihood principle \cite{Hradil2004,Paris2004}. 

In this work, we will consider three specifications. Firstly, our choice of estimation precision quantifier is the \emph{mean squared error} (MSE), $\Delta^2\thetahatN$, which is the expected squared difference between the true value of the parameter and its estimate. That is, 
\begin{eqnarray}\label{eq:MSEn}
\Delta^2\thetahatN:=\mathbb{E}[(\theta-\thetahatN(D))^2]=\sum_D L(D|\theta)(\theta-\thetahatN(D))^2,
\end{eqnarray} 
where $L(D|\theta)$ is the likelihood of observing the data $D$, given the true parameter value $\theta$. That is, for any $D=\{k_1,k_2,\cdots,k_\nu\}$ which contains $\nu_\ell$ occurrences of the outcome $\ell$,
\begin{eqnarray}\label{eq:defL} 
	L(D|\theta)= \prod_{\ell} p_\ell(\theta)^{\nu_\ell}, \quad\mathrm{with}\quad \sum_{\ell} \nu_{\ell}=\nu.
\end{eqnarray}
Secondly, we consider locally unbiased estimators, for which the MSE will be lower bounded by the \emph{Cram\'{e}r-Rao bound} (CRB)~\cite{Kay1993,Braunstein1994}: 
\begin{eqnarray}\label{eq:CRB}
\nu \Delta^2\thetahatN
\;\geq\;\frac{1}{\FNtheta}\;\geq\;\frac{1}{\FQNtheta}, 
\end{eqnarray}
where 
\begin{eqnarray}\label{eq:FI}\fl
\FNtheta=\FNtheta[\rhoN(\theta),\{M_\ell\}]:=\sum_\ell \frac{\dot{p}_\ell(\theta)^2}{p_\ell(\theta)}, \quad \mathrm{and} \quad \FQNtheta=\FQNtheta[\rhoN(\theta)]:=\max_{\{M_\ell\}}\FNtheta,
\end{eqnarray}
are respectively the \emph{Fisher information} (FI) and the \emph{quantum FI} (QFI)---the largest FI upon optimizing the choice of measurement, with $\dot{p}_\ell(\theta)=\partial_\theta p_\ell(\theta)$. Lastly, we here focus on \emph{asymptotic quantum inference}\cite{Hayashi2005}, where we study the ultimate precision obtainable in the limit of unlimited number of repetitions, $\nu\rightarrow\infty$. Realistically, we do not have unlimited repetitions of course, but the CRB will get sufficiently tight with large but finite $\nu$. Then, the inequalities in the CRB, Eq.~(\ref{eq:CRB}),  are saturated asymptotically\cite{Hayashi2005,Kay1993}, and we may turn our attention to the QFI as the figure of merit for the optimal attainable estimation precision. 

Given $\rhoN(\theta),$ the QFI and a choice of measurement that achieves it is well known:\cite{Helstrom1976, Paris2009}
\begin{eqnarray}\label{eq:QFI}
\FQNtheta[\rhoN(\theta)]=\tr\{\rhoN(\theta)L_\theta^2\},
\end{eqnarray}
where the Hermitian symmetric-logarithmic derivative (SLD) $L_\theta$ is defined implicitly by
\begin{eqnarray}\label{eq:SLD}
\partial_\theta\rhoN(\theta):=\frac{1}{2}\big\{L_\theta,\rhoN(\theta)\big\},
\end{eqnarray}
where $\big\{a,b\big\}=ab+ba$ is the anti-commutator. Moreover, the QFI can be attained by choosing the projective measurement into the eigenstates of the SLD, though in general cases there could be other possible choices as well.

As a remark, for $\rhoN(\theta)=\rho(\theta)^{\otimes N}$ having a product state structure, it follows from additivity of QFI \cite{Toth2014} that $\FQNtheta=N\mathcal{F}_{1;\theta}$, and the SLD will have a single-particle operator or an independent-sum structure, i.e., $L_\theta=\sum_{i=1}^N L_\theta^{(i)}$. Hence, \emph{local measurements} are sufficient to attain the QFI. This is expected and can be understood intuitively: with uncorrelated information carriers, we will never obtain more than the sum of optimal information from each individual. In this case then, the division into $\nu$ repetitions and $N$-ensemble can be seen as superfluous, as whatever the $\nu$ and $N$ separately are for fixed $\Ntot=\nu N\rightarrow\infty$, they all correspond to the same situation of having $\Ntot$ independent local measurements on $\Ntot$ independent qudits i.e., $\Ntot \Delta^2\hat{\theta}_N\geq \mathcal{F}_{1;\theta}^{-1}$. More generally though, it does matter how we divide the $\Ntot$ copies into $\nu$ repetitions of $N$-qudit ensemble, especially when we consider POM beyond local measurements on each qudits. Such a consideration is necessary, for example, when the local measurements are noisy due to some imperfect implementation, e.g., crosstalks mixing up the measurement outcomes locally, resulting in the effective POM elements no longer being rank-1 projectors. On one hand, insistence on using local measurements would thus have the achievable precision of estimating $\theta$ in $\rhoN(\theta)$ lower limited by $\Ntot\Delta^2\thetahatN\geq \lambda(\mathcal{F}_{1;\theta})^{-1}$ for some constant $\lambda>1$ that is characteristic of the noise. On the other hand, by allowing \emph{collective} or \emph{joint measurements}, interestingly, one can show that the measurement noise can be effectively negated, and now $\Ntot\Delta^2\thetahatN\geq c_N(\mathcal{F}_{1;\theta})^{-1}$, with an $N$-\emph{dependent} coefficient $c_N$ that $\rightarrow1$ in the large $N$ limit \cite{Len2021}. In this situation then, it pays off to have larger $N$, so long as $\nu$ is still sufficiently large to keep the CRB tight. In this work we shall not consider complications due to imperfect measurements, though as we will see next, we still nevertheless need to consider collective measurement for joint estimation of multiple parameters, and keep the distinction between $\nu$ repetitions and grouping into $N$-qudit ensemble for good.

\subsection{Joint-parameter estimation}\label{sec:joint-parameter estimation}
We move on to review the 
estimation of multiple parameters $\thetavec=\{\theta_1,\theta_2,\cdots,\theta_K\}$ encoded in the state $\rho(\thetavec)$. As in previous section, we consider $N$ identical copies of the qudit grouped together and described by the state $\rhoN(\thetavec)=\rho(\thetavec)^{\otimes N}$, and construct estimates of the parameters using asymptotic locally unbiased estimators $\thetahatNvec=\{\hat{\theta}_{N;1},\hat{\theta}_{N;2},\cdots,\hat{\theta}_{N;K}\}$ with the data $D$ collected from $\nu$ repetitions of a measurement of $\rhoN(\thetavec)$ with some $N$-qudit POM $\{M_\ell\}$. 
Then, the MSE is generalized to the \emph{covariance matrix}
$\cmatN$, whose $(i,j)$-th matrix element is given by
\begin{eqnarray}\label{eq:cmat} \fl
\mathcal{C}_{N;i,j}= \mathbb{E}[(\theta_i-\hat{\theta}_{N;i}(D))(\theta_j-\hat{\theta}_{N;j}(D))]=\sum_D L(D|\thetavec) (\theta_i-\hat{\theta}_{N;i}(D))(\theta_j-\hat{\theta}_{N;j}(D)),
\end{eqnarray}
where $p_\ell(\thetavec)=\tr\{\rhoN(\thetavec)M_\ell\}$ and $L(D|\thetavec)= \prod_{\ell} p_\ell(\thetavec)^{\nu_\ell}$ with $\sum_{\ell} \nu_{\ell}=\nu$, whereas the FI is generalized to the \emph{FI matrix} $F_{N;\thetavec}$, whose $(i,j)$-th matrix element reads
\begin{eqnarray}\label{eq:FIij}
F_{N;i,j}=\sum_\ell \frac{1}{p_\ell(\thetavec)}\frac{\partial p_\ell(\thetavec)}{\partial \theta_i}\frac{\partial p_\ell(\thetavec)}{\partial \theta_j}.
\end{eqnarray}
It then follows that we have the multi-parameter CRB, a matrix inequality, which reads\footnote{Note that in the literature, the multi-parameter CRB (as well as the Holevo and QFI bounds in Eq.~(\ref{eq:HolevoCRB})) is usually presented with $\nu=1$. In this case, each data $D$ is associated with a single outcome $\ell$, and the likelihood function $L(D|\thetavec)$ will be in one-to-one correspondence with $p_{\ell}(\thetavec)$, and we have $\mathbb{E}[\cdots]=\sum_{D}L(D|\thetavec)(\cdots)\rightarrow\sum_{\ell}p_{\ell}(\thetavec)(\cdots)$.}
\begin{eqnarray}\label{eq:mCRB}
\nu\cmatN \geq F_{N;\thetavec}^{-1}.  
\end{eqnarray}
Equivalently, Eq.~(\ref{eq:mCRB}) can be written as a numerical bound $\nu\tr\{\cmatN W\}\geq\tr\{F_{N;\thetavec}^{-1}W\}$ for any non-negative $K\times K$ matrix $W$, known as the cost matrix or weight matrix. Then, while there are a rich choices of possible bounds one may consider, such as the right logarithmic derivative (RLD) CRB \cite{Yuen1973}, for simplicity of presentation, here we mention just two: first, the covariance matrix can be further lower limited by the \emph{Holevo bound}, then by the \emph{multi-parameter QFI CRB}: \cite{Holevo1982, Nagaoka1989}
\begin{eqnarray}\label{eq:HolevoCRB}\fl
\nu\tr\{\cmatN W\}\geq\tr\{F_{N;\thetavec}^{-1}W\}\geq \min_{\{X_i\}}\{\tr\{W\Re Z\} + \|W\Im Z\|_1\}\geq \tr\{\FQNthetavec^{-1}W\}.
\end{eqnarray}
In the Holevo bound, $\|\cdot\|_1$ is the trace norm, $Z$ is the matrix with elements $Z_{i,j}=\tr\{X_i X_j\rhoN(\thetavec)\}$, where $\{X_i\}_{i=1}^K$ is a set of Hermitian matrices satisfying $\tr\{X_i\frac{\partial\rho_N(\thetavec)}{\partial\theta_j}\}=\delta_{i,j}$ and $\tr\{\rho_N(\thetavec)X_i\}=0\,\forall i$. Meanwhile, in the multi-parameter QFI CRB, we have the \emph{QFI matrix} $\FQNthetavec$, with matrix elements $\mathcal{F}_{N;i,j}:=\frac{1}{2}\tr\{\rhoN(\thetavec)\{L_{\theta_i},L_{\theta_j}\}\}$, where the SLDs are defined analogously as in Eq.~(\ref{eq:SLD}), i.e., $\partial_{\theta_i}\rhoN(\thetavec):=\frac{1}{2}\big\{L_{\theta_i},\rhoN(\thetavec)\big\}$. Note that the diagonal elements of the QFI matrix are exactly the QFI for the individual parameters as in single-parameter estimation, i.e., $\mathcal{F}_{N;i,i}=\mathcal{F}_{N;\theta_{i}}$, while the diagonal elements of the covariance matrix, are, of course, the MSE for the individual parameters, i.e., $\nu\mathcal{C}_{N;i,i}=\nu\Delta^2\hat{\theta}_{i;N}$. Moreover, both the Holevo bound and QFI matrix is additive, i.e., the Holevo bound for $\rhoN(\thetavec)=\rho(\thetavec)^{\otimes N}$ is $N^{-1}$ times of that for $\rho(\thetavec)$ \cite{Hayashi2008,Yamagata2013}, and $\FQNthetavec=N\mathcal{F}_{1;\thetavec}$. 

By the same reasoning as in the single-parameter cases, the first inequality in Eq.~(\ref{eq:HolevoCRB}) can always be saturated \cite{Kay1993} in the large repetition limit, $\nu\rightarrow\infty$. Remarkably then, in the large ensemble limit $N\rightarrow\infty$, up to the first order convergence rate, i.e., the leading order of $1/N$, the Holevo bound can always be attained as well (with some regularity conditions satisfied)\cite{Hayashi2008,Guta2006,Kahn2009,Yamagata2013,Yang2019b}. The last, weaker multi-parameter QFI CRB however, is only tight (asymptotically) and equal to the Holevo bound for asymptotically classical model \cite{Suzuki2019}, where the \emph{weak commutativity} condition $\tr\{\rhoN(\thetavec)[L_{\theta_i},L_{\theta_j}]\}=0\,\forall i\neq j$ is satisfied, where $[a,b]=ab-ba$ is the commutator\cite{Matsumoto2002,Vaneph2013,Crowley2014,Ragy2016}. Finally, should a more stringent condition $\tr\{\rhoN(\thetavec)L_{\theta_i}L_{\theta_j}\}=0\,\forall i\neq j$ be met, we have $\mathcal{F}_{N;i,j}=0 \,\forall i\neq j$, and the QFI matrix is diagonal. In this case then, the inverse of the QFI matrix can be done element-wise, and we have $\nu\mathcal{C}_{N;i,i}=\nu\Delta^2\hat{\theta}_{i;N}=1/\mathcal{F}_{N;i,i}=1/\mathcal{F}_{N;\theta_{i}}$ as $N\rightarrow\infty$, i.e., for all the parameters, there exist a measurement scheme which allows us to estimate them simultaneously, each with an optimal asymptotic precision that is equal to that of the corresponding single-parameter estimation scenario. We shall refer to this as ``compatible" joint-parameter estimation, as in this case we can optimally estimate one particular parameter without affecting the other at all. As a remark, note that since the QFI matrix is non-negative, it can in fact be always diagonalized in a certain basis---equivalently, we can always define a new set of parameters $\thetavec'(\thetavec)$ from the initial parameters $\thetavec$, where now their joint estimation is compatible. In the statistic literature, the subject of diagonalizing a FI matrix is also known as \emph{parameter orthogonality} \cite{Cox1987}, and has also been discussed in the context of nuisance parameters in quantum estimation \cite{Suzuki2020}, where only a single parameter is of actual interest, and hence a reparametrization might be beneficial in constructing a good estimator for it. In general though, the new parameters $\thetavec'$ might change their form for different true values of $\thetavec$, as the diagonalization of the QFI matrix depends locally on $\thetavec$. Here, we are concern with global parameter orthogonality, i.e., joint estimation of some given, fixed parameters $\thetavec$ for different possible true values, in which $\tr\{\rhoN(\thetavec)L_{\theta_i}L_{\theta_j}\}=0\,\forall i\neq j$ is then a sufficient condition their compatibility. In this sense, our definition of compatibility here follows that of Ref.~\cite{Ragy2016}, which is different from those for example in Refs~\cite{Sidhu2021, Lu2021, Belliardo2021, Miyazaki2021}, where compatibility is defined for saturation of the multi-parameter QFI CRB even if the QFI matrix is non-diagonal globally for all parameter values.

\section{Two-parameter estimation for qubit states}\label{Sec3}
In this section, we apply the general formalism described above to the estimation of the length and direction of Bloch vector for a qubit state. Then, we show how one can asymptotically achieve compatible joint estimation for the two parameters, by a measuring an observable that can be thought of as the total angular momentum squared operator, followed by a commuting, local projective measurement. Furthermore, we look deeper into the case of nearly-pure qubit estimation, and reveal a connection with the problem of superresolution imaging in the sub-Rayleigh limit.

\subsection{The qubit model}\label{sec:qubitmodel}

\begin{figure*}[t!]
	\centering
	\includegraphics[width=0.4\textwidth]{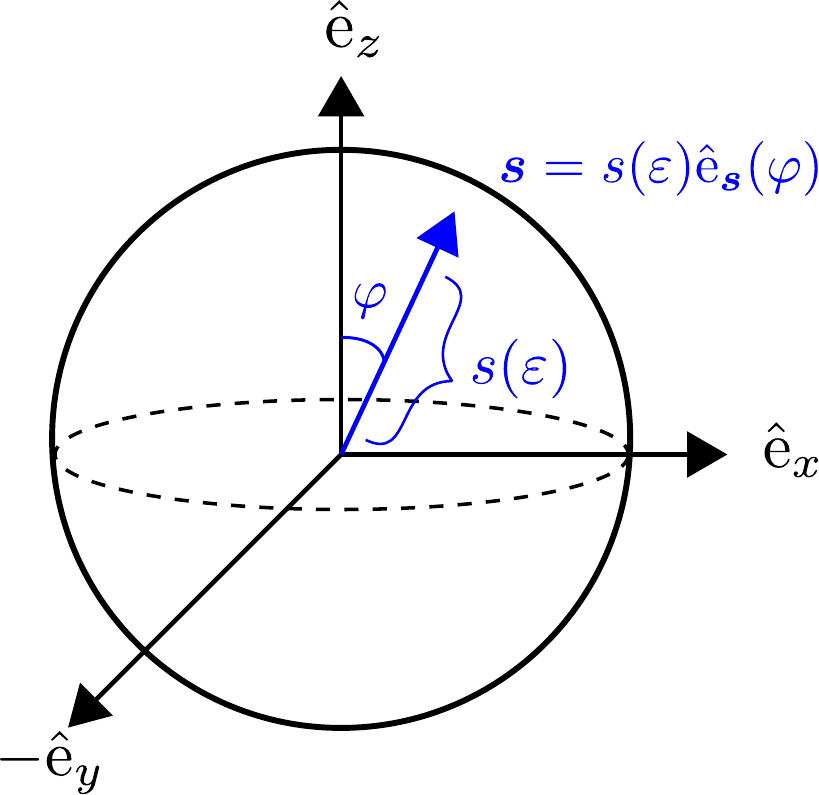}
	\caption{\textbf{Bloch-sphere representation of the qubit model}. The qubit state is represented by the Bloch vector $\boldsymbol{s}$, with its length given by $s(\varepsilon)$ and direction $\hat{\upe}_{\boldsymbol{s}}(\varphi)=\sin\varphi~\hat{\upe}_{x}+\cos\varphi~\hat{\upe}_{z}$. The parameter of interest of estimation are $\varphi$ and $\varepsilon$.}
	\label{fig:MPEfQS-fig1}
\end{figure*}

Consider a qubit described by the state $\rho(\boldsymbol{\theta})$, which in the familiar Bloch-sphere representation as visualized in Fig.~\ref{fig:MPEfQS-fig1}, has a Bloch vector $\boldsymbol{s}$ that is confined to the $x$-$z$ plane, and deviates from the $z$-axis by some angle $\varphi$. We allow the length of the Bloch vector to be possibly further parametrized by the variable $\varepsilon$, such that 
\begin{eqnarray}\label{eq:BlochvecS}
\fl
\rho(\boldsymbol{\theta})=\rho(\varepsilon,\varphi)=\frac{1}{2}\big(\id+\boldsymbol{s}\cdot\boldsymbol{\sigma}\big), \quad \boldsymbol{s}=s(\varepsilon)\hat{\upe}_{\boldsymbol{s}}(\varphi), \quad \hat{\upe}_{\boldsymbol{s}}(\varphi)=\sin\varphi~\hat{\upe}_{x}+\cos\varphi~\hat{\upe}_{z},
\end{eqnarray}
where $\boldsymbol{\sigma}$ is the usual Pauli operator vector. The two parameters of interest of estimation are $\thetavec=\{\varepsilon, \varphi\}$. Note that in Refs.~\cite{Hayashi2008, Bagan2006} which study more general qubit model with up to three parameters, our qubit model here has also been studied as a sub-model with just two parameters, mainly from the prospect of analyzing the Holevo bound.

It is helpful to think of the qubit as a spin-1/2 system, and so equivalently  $\rho(\boldsymbol{\theta})$ can be expressed as
\begin{eqnarray} \label{eq:rhoSingle}
\rho(\varepsilon,\varphi)&=\upe^{-\upi \sigma_y\varphi/2}\frac{\upe^{\beta \sigma_z}}{2\cosh\beta}\upe^{\upi \sigma_y\varphi/2},
\end{eqnarray}
where $\tanh\beta=s(\varepsilon)$ and $\sigma_\ell$, $\ell={x,y,z}$, is the  Pauli-$\ell$ operator. With Eq.~(\ref{eq:rhoSingle}), $N$ identical copies of the qubit in $\rho(\boldsymbol{\theta})$ can then be treated as a collection of spin-1/2s, with the $N$-qubit state written concisely as
\begin{eqnarray}\label{eq:rhoN}
\rhoN(\boldsymbol{\theta})=\rhoN(\varepsilon,\varphi)=\rho(\varepsilon,\varphi)^{\otimes N}=\upe^{-\upi J_y\varphi}\frac{\upe^{2\beta J_z}}{(2\cosh\beta)^N}\upe^{\upi J_y\varphi},
\end{eqnarray}
with the generalization to $N$-spin angular momentum, i.e., $\Jvec=\frac{1}{2}\sum_{i=1}^N \boldsymbol{\sigma}^{(i)}$, where $\boldsymbol{\sigma}^{(i)}$ is the Pauli operator vector for the $i$-th qubit. Note that, as can be seen immediately from Eqs.~(\ref{eq:rhoSingle}, \ref{eq:rhoN}), we can also treat our state parameter estimation problem as a quantum metrology or channel parameter estimation problem, where the state $\rhoN(\thetavec)$ is obtained by encoding the parameters $\thetavec$ into an input product state $\rho_{0}$ polarized in the $+z$ direction, using $N$-independent unitary channel $\mathcal{U}_\varphi\sim\{U_\varphi=\upe^{\upi\sigma_y \varphi/2}\}$ that each suffers a dephasing noise--- $\Lambda_\eta\sim \{\sqrt{(1+\eta)/2}\id,\sqrt{(1-\eta)/2}\sigma_y\}$ with $\eta=s(\varepsilon)$. That is, we can equivalently think of $\rhoN(\thetavec)$ in Eq.~(\ref{eq:rhoN}) as
\begin{eqnarray}\label{eq:QMmodel}
\rho_N(\thetavec)=\Lambda_\eta^{\otimes N}\circ \mathcal{U}_\varphi^{\otimes N}[\rho_{0}], \quad \rho_{0}=\Big(\frac{1}{2}(\id+\sigma_z)\Big)^{\otimes N}.
\end{eqnarray}
In this work, we will work on estimation of $\thetavec$ in the state $\rhoN(\thetavec)$ as given in Eq.~(\ref{eq:rhoN}) regardless of its interpretation. Nevertheless, we will return for a short discussion in Sec.~\ref{Sec5} on comparison of estimation precision in the corresponding quantum metrology scenario, where instead of using just product input states for $\rho_0$, one may consider the use of entangled states as well.

\subsection{Single-parameter estimation}
First, let us work out the estimation precision limit, should we just estimate one of the parameter in the state $\rhoN(\varepsilon,\varphi)$ in Eq.~(\ref{eq:rhoN}), with another parameter being actually known. 

\subsubsection{Estimation of $\varepsilon$}
With $\varphi$ being known and fixed, it is straightforward to find that 
\begin{eqnarray}\label{eq:SLDeps}
L_\varepsilon&=\frac{1}{1-s(\varepsilon)^2}\big[2\upe^{-\upi J_y\varphi}J_z\upe^{\upi J_y\varphi}-Ns(\varepsilon)\id\big] \frac{\partial s(\varepsilon)}{\partial\varepsilon},
\end{eqnarray}
and 
\begin{eqnarray}\label{eq:QFIeps}
\FQNeps&=
N\frac{1}{1-s(\varepsilon)^2}\Big(\frac{\partial s(\varepsilon)}{\partial\varepsilon}\Big)^2.
\end{eqnarray}
Then, as $\upe^{-\upi \sigma_y\varphi/2}\sigma_z\upe^{\upi \sigma_y\varphi/2}=\hat{\upe}_{\boldsymbol{s}}\cdot\boldsymbol{\sigma}$, from Eq.~(\ref{eq:SLDeps}), the optimal precision $\FQNeps$ can be attained by performing the projective measurement specified by the directions $\pm\hat{\upe}_{\boldsymbol{s}}$, i.e., $\{\frac{\id+\hat{\upe}_{\boldsymbol{s}}\cdot\boldsymbol{\sigma}}{2},\frac{\id-\hat{\upe}_{\boldsymbol{s}}\cdot\boldsymbol{\sigma}}{2}\}$ for each of the qubit, which has the respective probability $p_\pm(\varepsilon)=\frac{1\pm s(\varepsilon)}{2}$ of obtaining them, and so $\varepsilon=s^{-1}(p_+(\varepsilon)-p_-(\varepsilon))$. Note that in this case the optimal measurement does not depend on the parameter $\varepsilon$ itself.

An unbiased estimator (in the asymptotic $\nu\rightarrow\infty$ limit) $\hat{\varepsilon}_N$ that attains the QFI $\FQNeps$ for any $N$ is the maximum-likelihood estimator (MLE) \cite{Hradil2004,Paris2004}. More precisely, denote $\nu_{\pm}^{(i)}$ as the relative frequencies of obtaining the outcome of  $\frac{\id\pm\hat{\upe}_{\boldsymbol{s}}\cdot\boldsymbol{\sigma}^{(i)}}{2}$ for the $i$-th qubit over the $\nu$ repetitions, such that $\nu_+^{(i)}+\nu_-^{(i)}=1$ for all $i$. Then, our data is essentially $D=\{\nu_+,\nu_-\}$ with $\nu_{\pm}:=\sum_{i=1}^N\frac{\nu_{\pm}^{(i)}}{N}$, and the MLE is defined by
\begin{eqnarray}\label{eq:MLEvareps}
	\hat{\varepsilon}_N(D):= \arg \max_\varepsilon L(D|\varepsilon)&=&\arg \max_\varepsilon p_+(\varepsilon)^{N\nu\nu_+}p_-(\varepsilon)^{N\nu\nu_-}.
\end{eqnarray}
Note that, for data where $\nu_+-\nu_-$ is non-negative (which will have overall likelihood approaching one as $\nu\rightarrow\infty$ by virtue of the central limit theorem such that $\sqrt{\nu}\nu_\pm$ converges to a Gaussian peaked at $\sqrt{\nu}p_\pm(\varepsilon)$), the MLE can be conveniently written as $\hat{\varepsilon}_N(D):=s^{-1}(\nu_+-\nu_-)$, which coincides with the linear inversion estimator. Otherwise, the maximization in Eq.~(\ref{eq:MLEvareps}) needs to be carried out explicitly, and the MLE will in general have a small bias (which vanishes as $\nu\rightarrow\infty$) as compared to the linear inversion estimator.

\subsubsection{Estimation of $\varphi$}\label{sec:varphi}
With $\varepsilon$ being known and fixed instead, we find that
\begin{eqnarray}
L_\varphi&=2s(\varepsilon)\upe^{-\upi J_y\varphi}J_x\upe^{\upi J_y\varphi},  \label{eq:SLDphi} \\
\FQNphi&=Ns(\varepsilon)^2, \label{eq:QFIphi} 
\end{eqnarray}
where the optimal precision $\FQNphi$ can be obtained with the projective measurement onto the direction specified by $\pm\hat{\upe}_{\boldsymbol{s}'}$ for each qubit, where $\hat{\upe}_{\boldsymbol{s}'}=\cos\varphi\,\hat{\upe}_{x}-\sin\varphi\,\hat{\upe}_{z}$. Note that in this case the optimal measurement does depend on the parameter $\varphi$ itself---a common feature in local estimation problem. In particular then, suppose we are performing measurement in the basis specified by $\pm\hat{\upe}_{\boldsymbol{s}'}$ with $\hat{\upe}_{\boldsymbol{s}'}=\cos\alpha\,\hat{\upe}_{x}-\sin\alpha\,\hat{\upe}_{z}$ where eventually $\alpha\rightarrow\varphi$, such that the respective probablity is given by $p_\pm(\varphi)=\frac{1\pm s(\varepsilon)\sin(\varphi-\alpha)}{2}$, and then $\varphi=\alpha+\sin^{-1} (\frac{p_+(\varphi)-p_-(\varphi)}{s(\varepsilon)})$. In order to attain the QFI $\FQNphi$ in the asymptotic limit $\nu\rightarrow\infty$ for any $N$ we may consider the MLE as defined by
\begin{eqnarray}\label{eq:MLEvarphi}
	\hat{\varphi}_N(D):= \arg \max_\varphi L(D|\varphi)&=&\arg \max_\varphi p_+(\varphi)^{N\nu\nu_+}p_-(\varphi)^{N\nu\nu_-},
\end{eqnarray}
where now $\nu_{\pm}:=\sum_{i=1}^N\frac{\nu_{\pm}^{(i)}}{N}$ is the sum of the relative frequencies for the projection onto $\frac{\id\pm\hat{\upe}_{\boldsymbol{s}'}\cdot\boldsymbol{\sigma}^{(i)}}{2}$. Similar to the discussion earlier, the MLE will coincide with the linear inversion estimator with $\hat{\varphi}_N(D):=\varphi+\sin^{-1} (\frac{\nu_+-\nu_-}{s(\varepsilon)})$ after taking $\alpha\rightarrow\varphi$, when $\frac{\nu_+-\nu_-}{s(\varepsilon)}$ is bounded between zero and one, which will be mostly the case as $\nu\rightarrow\infty$. As a remark, $\FQNphi$ can only be attained by the said projective measurement onto the direction $\pm\hat{\upe}_{\boldsymbol{s}'}$ with $\alpha=\varphi$ for any $s(\varepsilon)$ that is strictly not equal to zero or unity; for $s(\varepsilon)=1$, arbitrary values of $\alpha$ would also attain $\FQNphi$.

\subsection{Joint-parameter estimation}
First, from a quick observation where the state $\rhoN(\thetavec)$ (\ref{eq:rhoN}) and the two SLDs (\ref{eq:SLDeps}, \ref{eq:SLDphi}) are essentially all real in the same basis, we can immediately conclude that the weak commutativity condition holds. It then follows that the Holevo bound and the multi-parameter QFI CRB are equal, which dictate the first order convergence rate of the weighted trace of the covariance matrix as $N\rightarrow\infty$. Moreover, one can readily verify that we have $\tr\{\rhoN(\thetavec)L_\varepsilon L_\varphi\}=\tr\{\rhoN(\thetavec)L_\varphi L_\varepsilon\}=0$, and therefore the QFI matrix is diagonal, such that the two parameters are globally orthogonal and it is in principle possible to realize compatible joint estimation for them. That is, the optimal precisions in simultaneously estimating the two parameters in $\rhoN(\varepsilon,\varphi)$ are given by their respective QFI as in the single-parameter scenarios; see Eq.~(\ref{eq:QFIeps}) and Eq.~(\ref{eq:QFIphi}).  

Notice that, however, $L_\varepsilon L_\varphi \neq L_\varphi L_\varepsilon$, which means that the optimal local measurement bases for estimation of $\varepsilon$ and $\varphi$ actually do not commute. In fact, as $\hat{\upe}_{\boldsymbol{s}}\cdot\hat{\upe}_{\boldsymbol{s}'}=0$, they correspond to mutually unbiased bases \cite{Schwinger1960,Durt2010}---acquiring optimal information about one parameter as such completely erases information about the other, and therefore we could never obtain simultaneous estimation of both parameters with optimal precision by measuring the $N$ qubits separately. Therefore in order to attain still the multi-parameter QFI CRB, we must now consider collective or global measurements on the whole $N$-qubit ensemble instead.
 
\subsubsection{Measurement with $\Jvec^2$ and $L_\varphi$}
We look for collective measurements that preserve information about the second parameter, say $\varphi$, when accessing information about the first parameter, say $\varepsilon$. Then, as $\varphi$ serves as the angle of rotation generated by angular momentum $J_y$, see Eq.~(\ref{eq:rhoN}), it is inviting for us to consider measuring the angular momentum squared operator, $\Jvec^2$, which is invariant under rotation, for estimation of $\varepsilon$. Indeed, as the eigenvalues or spectrum of the state $\rho(\thetavec)$, which is $\frac{1\pm s(\varepsilon)}{2}$, is related exclusively to the parameter $\varepsilon$ but not the parameter $\varphi$, the estimation of $\varepsilon$ can also be viewed as estimation of the spectrum of $\rho(\thetavec)$, where Keyl and Werner have shown that the $\Jvec^2$ measurement on $\rhoN(\thetavec)$ produces estimates that have vanishing error as $N\rightarrow\infty$ \cite{Keyl2001}. This fact has also been discussed in \cite{Demkowicz2020}, and similar conclusions can also be observed in Refs.~\cite{Hayashi2008,Bagan2006}.

In spite of the insights and conclusions from the above references, let us here provide a slightly different but perhaps the most direct and explicit account on the MSE for estimation of $\varepsilon$ with the $\Jvec^2$ measurement, by applying the well-known error-propagation formula\cite{Wineland1992} for $\nu\rightarrow\infty$, which reads (see \ref{app:errorpropJvec2} for details of the evaluation)
\begin{eqnarray}\label{eq:J2errpro}
\nu\Delta^2\hat{\varepsilon}_N&= \frac{\tr\{(\Jvec^2)^2\rhoN(\thetavec)\}-\tr\{\Jvec^2\rhoN(\thetavec)\}^2}{|\frac{\partial}{\partial\varepsilon}\tr\{\Jvec^2\rhoN(\thetavec)\}|^2}\nonumber\\
&=\frac{1}{N}\Bigg[\frac{(3-2N)s(\varepsilon)^4+2(N-3)s(\varepsilon)^2+3}{2(N-1)s(\varepsilon)^2\left(\frac{\partial s(\varepsilon)}{\partial\varepsilon}\right)^2}\Bigg].
\end{eqnarray}
Eq.~(\ref{eq:J2errpro}) applies to most estimator $\hat{\varepsilon}_N$ that takes in the mean of $\Jvec^2$ (as estimated from the data by $\langle\Jvec^2\rangle(D)=\sum_{j=N/2-\lfloor N/2 \rfloor}^{N/2} j(j+1)\nu_j$, where $\nu_j$ is the relative frequency for the projection onto the $j$-subspace of $\Jvec^2$ over $\nu$ repetitions) as input, and precisely estimates $\varepsilon$ when the estimated mean is exactly the true mean value, i.e., $\langle\Jvec^2\rangle(D)=\Tr\{\rhoN(\thetavec)\Jvec^2\}$. Then, we check that
in the large $N$ limit, $\nu\Delta^2\hat{\varepsilon}_N$ has the asymptotic behaviour
\begin{eqnarray}\label{eq:J2errprolargeN}
\nu\Delta^2\hat{\varepsilon}_N &=\frac{1}{N}\Big[\frac{N}{N-1}(1-s(\varepsilon)^2)+\frac{3s(\varepsilon)^2-6+3s(\varepsilon)^{-2}}{2(N-1)}\Big]\left(\frac{\partial s(\varepsilon)}{\partial\varepsilon}\right)^{-2}  \nonumber \\
&\sim \frac{1}{N}(1-s(\varepsilon)^2)\left(\frac{\partial s(\varepsilon)}{\partial\varepsilon}\right)^{-2}+\Or\Big(\frac{1}{N^2}\Big)\nonumber\\
&=\FQNeps^{-1}+\Or\Big(\frac{1}{N^2}\Big),
\end{eqnarray} 
where $\Or(1/N^2)$ stands for terms that are of order $1/N^2$ and beyond. We thus have a clear demonstration that as $N$ gets ever larger, $\nu\Delta^2\hat{\varepsilon}_N\sim \frac{N}{N-1}\FQNeps^{-1}\rightarrow \FQNeps^{-1}$, i.e., indeed the $\Jvec^2$ measurement permits construction of estimator which asymptotically estimates $\varepsilon$ as precise as using the optimal projective measurement when assuming the parameter $\varphi$ is known (c.f. Eqs.~(\ref{eq:SLDeps}, \ref{eq:QFIeps})). Importantly, although $\nu\Delta^2\hat{\varepsilon}_N\rightarrow \FQNeps^{-1}$ only asymptotically, the estimation of $\varepsilon$ is global here, in the sense that the observable $\Jvec^2$ is independent of the two parameters $\varepsilon$ and $\varepsilon$, and so allows us to continue estimating $\varphi$ without any potential lost of information as desired. That is, as $\Jvec^2$ commutes with $L_\varphi$---one can explicitly confirm with Eq.~(\ref{eq:SLDphi}), we may still effectively measure in the $L_\varphi$ basis, and obtain, in the $\nu\rightarrow\infty$ limit for any $N$,
\begin{eqnarray}\label{eq:MSEvarphi}
\nu\Delta^2\hat{\varphi}_N = \FQNphi^{-1},
\end{eqnarray}
using for example the MLE $\hat{\varphi}_N$ introduced in Eq.~(\ref{eq:MLEvarphi}) in Sec.~\ref{sec:varphi}. 

\begin{figure*}[t!]
	\centering
	\includegraphics[width=0.7\textwidth]{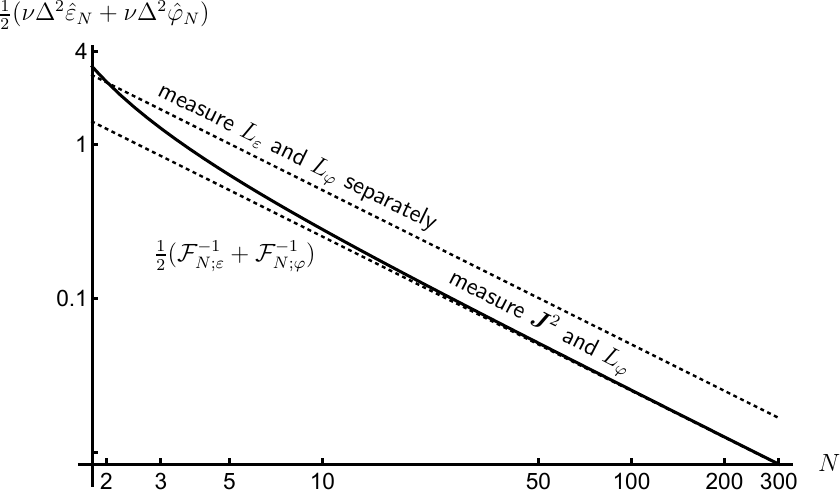}
		\caption{\textbf{Weighted sum of mean squared error (MSE) in estimating $\varepsilon$ and $\varphi$ using the global $\Jvec^2$ and local $L_\varphi$ measurement} as given by Eq.~(\ref{eq:MSEvarphi}) and Eq.~(\ref{eq:J2errpro}), for different $N$ values (\emph{thick solid line}). The lower \emph{dashed line} corresponds to the ultimate (weighted) achievable precision should we have perfect knowledge about the other parameter, and only a single variable to estimate (which for our model, is also the Holevo bound). Note that the plot starts at $N=2$, as the classification of collective versus local measurements is not meaningful for $N=1$, and in this plot we consider a specific parametrization of $s(\varepsilon)=1-\varepsilon^2/8$ with $\varepsilon^2/8=0.1$, though the convergence behaviour is general and independent of the parametrization. For comparison, we also include the best (weighted) MSE obtainable (upper \emph{dashed line}), should we use the strategy of dividing equally the $\nu$ resources into estimating $\varepsilon$ and $\varphi$ separately. The advantage of $\Jvec^2$ collective measurement is evident for $N\geq3$.}
		\label{fig:MPEfQS-fig2}
\end{figure*}

In Fig.~\ref{fig:MPEfQS-fig2}, we illustrate the convergence of the weighted sum of two the estimation precision, $\frac{1}{2}\nu\Delta^2\hat{\varepsilon}_N+\frac{1}{2}\nu\Delta^2\hat{\varphi}_N$, with the $\Jvec^2$ and $L_\varphi$ measurement strategy, to the ultimate bound $\frac{1}{2}\FQNeps^{-1}+\frac{1}{2}\FQNphi^{-1}$, for the chosen parametrization of $s(\varepsilon)=1-\varepsilon^2/8$. 

\subsubsection{Nearly-pure qubit estimation}\label{sec:nearly-pure}
Having established the general results for compatible joint estimation for $\varepsilon$ and $\varphi$, let us look deeper into the scenario where the qubits are nearly pure, i.e., $s(\varepsilon)\approx1$. In particular, we focus on ``regular" parametrization where $s(\varepsilon)$ is smooth around some value $\varepsilon_0$ with $s(\varepsilon_0)=1$, such that we can perform the Taylor expansion in powers of $\delta\ll1$ around $\varepsilon_0$ for some $\varepsilon=\varepsilon_0-\delta$. Then, while the conclusion below is generally true for any regular parametrization of $s(\varepsilon)$ as well, for explicit illustration, let us consider specifically $s(\varepsilon)=1-\varepsilon^2/8$ with $\varepsilon\ll1$, a choice that is motivated by a case study in quantum imaging superresolution problem: As reported in Ref.~\cite{Chrostowski2017}, estimating $\varphi$ and $\varepsilon$ in this parameterization could be seen as estimating the centroid and the separation between two close incoherent light sources (up to some multiplicative constant), respectively. 

\begin{figure*}[t!]
	\centering
	\includegraphics[width=0.81\textwidth]{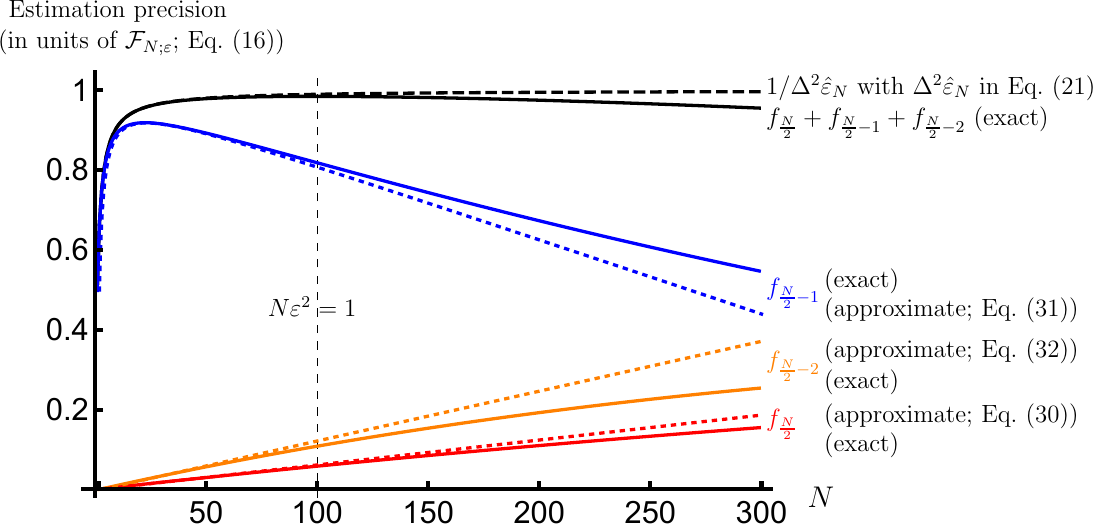}
	\caption{\textbf{Fisher information (FI) of the largest three angular momentum modes} for estimation of $\varepsilon$ for different $N$ values, with the parametrization $s(\varepsilon)=1-\varepsilon^2/8$ with $\varepsilon^2=0.01$. The \emph{red lines} are the FI as contributed by the largest $j=N/2$ mode, with solid line for the exact values, and dotted line for the approximate values as given by Eq.~(\ref{eq:fN/2}). Similarly, the \emph{blue and orange lines (solid for exact, dotted for approximate)} correspond respectively for the mode $j=N/2-1$ and $j=N/2-2$. Their (exact) joint contribution is depicted by the \emph{black solid line}, which we compare against with the whole information obtained from the $\Jvec^2$ measurement (\emph{black dashed line})---essentially the inverse of $\Delta^2\hat{\varepsilon}_N$ in Eq.~(\ref{eq:J2errpro}). Despite up to $N=300$ in this case where the three modes still account for most of the information obtained with $\Jvec^2$ measurement, we see that it is decreasing as $N\varepsilon^2$ starts to get larger than one, which is a manifestation of the fact that information carried in the lower angular momentum modes are now becoming significant.}
	\label{fig:MPEfQS-fig3}
\end{figure*}

We shed some light as to how the $\Jvec^2$ measurement works in this $\varepsilon\ll1$ regime. From the theory of angular momentum, see textbooks like Refs.~\cite{CohenTannoudji, Gottfried2013}, we know that the full $N$-spin Hilbert space, $\mathcal{H}_\mathrm{total}$, can be decomposed into direct sum of orthogonal subspaces specified by two numbers, $j$ and $g(j)$:
\begin{eqnarray}\label{eq:Htotal}
\mathcal{H}_\mathrm{total}&= \Moplus_{j=N/2-\lfloor N/2 \rfloor}^{N/2} \Big(\Moplus_{g(j)=1}^{\mu_j}  \mathcal{H}_{j,g(j)}\Big). 
\end{eqnarray}
Here, $j=\{N/2, N/2-1, \ldots, N/2-\lfloor N/2 \rfloor\}$ is nothing but the usual quantum number associated with the total angular momentum squared, while $g(j)=\{1,\ldots,\mu_j-1,\mu_j\}$ specifies the $\mu_j$ different subspaces for the same $j$ value. Generally, the multiplicity factor $\mu_j$ depends on how the total angular momentum is composed; for our case of $N$ spin-1/2s, it is known that, see for example Refs.~\cite{Cirac1999} and \cite{Banaszek2001} \footnote{There is a typo in the equation (A4) of Ref.~\cite{Banaszek2001}.}, 
$\mu_j=\binom{N}{N/2-j}-\binom{N}{N/2-j-1}=\frac{2j+1}{N+1}\binom{N+1}{N/2-j}$.
\noindent Denote $\cP_j$ as the projector onto the $j$-angular momentum space, i.e., the space $\Moplus_{g(j)=1}^{\mu_j}  \mathcal{H}_{j,g(j)}$. Then, with the quantum state $\rhoN(\varepsilon,\varphi)$ of Eq.~(\ref{eq:rhoN}), evidently $\rhoN(\varepsilon,\varphi)=\sum_{j=N/2-\lfloor N/2\rfloor}^{N/2}\cP_j \rhoN(\varepsilon,\varphi) \cP_j$, and the probability that the $N$-qubit ensemble is found to have the total angular momentum value $j$ is given by 
\begin{eqnarray}\label{eq:pj}
p_j&=\tr\Big\{\cP_j \upe^{-\upi J_y\varphi}\upe^{2\beta J_z}\upe^{\upi J_y\varphi}\Big\}/(2\cosh\beta)^N=\mu_j \sum_{m=-j}^{j}  \upe^{2\beta m}/(2\cosh\beta)^N\nonumber\\
&=\mu_j \frac{\sinh\big(\beta(1+2j)\big)}{\sinh\beta}\frac{1}{(2\cosh\beta)^N}\nonumber\\
&= \mu_j \frac{\big[\varepsilon\sqrt{16-\varepsilon^2}\big]^{-2j}\big[(16-\varepsilon^2)^{1+2j}-(\varepsilon^2)^{1+2j}\big]}{16(1-\varepsilon^2/8)}\frac{\varepsilon^N}{4^N}\Big(1-\frac{\varepsilon^2}{16}\Big)^{N/2}. 
\end{eqnarray}
Notice that according to expectations $p_j$ does not depend on $\varphi$. As $\varepsilon\ll1$, we expand Eq.(\ref{eq:pj}) in powers of $\varepsilon$, and the first two leading terms are 
\begin{eqnarray}\label{eq:pj2nd}
p_j&\approx\mu_j \Big(\frac{\varepsilon}{4}\Big)^{N-2j}\Big[1+a_{j,N}\varepsilon^2\Big],
\end{eqnarray}
where $a_{j,N}=\frac{1}{16}(1-j-\frac{N}{2}-\delta_{j,0})$. More precisely, as the leading order of the series increases in powers of $N\varepsilon^2$, we see that the truncated series expansion in Eq.~(\ref{eq:pj2nd}) should approximate Eq.~(\ref{eq:pj}) sufficiently well if $N\varepsilon^2\ll1$. Then, from Eq.~(\ref{eq:pj2nd}) we see that the leading contribution for the FI, up to $N\varepsilon^2$, is from the three subspaces with $j=N/2, N/2-1$, and $N/2-2$. That is, 
\begin{eqnarray}
p_{\frac{N}{2}}&\approx1-\frac{N-1}{16}\varepsilon^2,\label{eq:pN/2}\\
p_{\frac{N}{2}-1}&\approx(N-1)\frac{\varepsilon^2}{16}\Big[1+\frac{(2-N)}{16}\varepsilon^2\Big], \quad [\mathrm{defined\, only\, for} N\geq2],\label{eq:pN/2-1}\\
p_{\frac{N}{2}-2}&\approx\frac{N(N-3)\varepsilon^4}{512}, \quad [\mathrm{defined\, only\, for} N\geq4], \label{eq:pN/2-2}
\end{eqnarray}
and with $f_j=\frac{1}{p_j} \Big(\frac{\partial p_j}{\partial\varepsilon}\Big)^2$, we have 
\begin{eqnarray}
f_{\frac{N}{2}}&\approx\frac{(N-1)^2}{64}\varepsilon^2, \label{eq:fN/2}\\
f_{\frac{N}{2}-1}&\approx\frac{N-1}{4}+\frac{3(N-1)}{64}\Big(2-N-\delta_{N,2}\Big)\varepsilon^2, \label{eq:fN/2-1}\\
f_{\frac{N}{2}-2}&\approx\frac{N(N-3)}{32}\varepsilon^2, \label{eq:fN/2-2}
\end{eqnarray}
such that
\begin{eqnarray}\label{eqFne2nd} 
\fl
F_{N;\varepsilon}=\sum_{j=N/2-\lfloor N/2\rfloor}^{N/2} f_j=f_{\frac{N}{2}}+f_{\frac{N}{2}-1}+f_{\frac{N}{2}-2}+ \big[\textsf{terms of $\Or(\varepsilon^4)$ and higher} \big]\nonumber \\
\fl \hphantom{F_{N;\varepsilon}}\approx \frac{N-1}{4}+\Big[\frac{(N-1)^2}{64}+\frac{3(N-1)}{64}(2-N-\delta_{N,2})+\frac{N(N-3)}{32}\eta(N-4)\Big]\varepsilon^2, 
\end{eqnarray}
where $\eta(\cdot)$ is the unit step function, and here we define $\eta(0)\equiv1$. For $N\geq4$, the FI up to $N\varepsilon^2$ is then equal to
\begin{eqnarray}
F_{N;\varepsilon}&=\frac{N}{4}\Big(1+\frac{\varepsilon^2}{16}\Big)-\Big(\frac{1}{4}+\frac{5\varepsilon^2}{64}\Big).
\end{eqnarray}
Upon comparing with the QFI in Eq.~(\ref{eq:QFIeps}) that now reads
\begin{eqnarray}\label{eq:FQNeps-exp}
\FQNeps=N\frac{4}{16-\varepsilon^2}=\frac{N}{4}\Big(1+\frac{\varepsilon^2}{16}+\frac{\varepsilon^4}{256}+\cdots\Big),
\end{eqnarray}  
we then confirm that in the large $N$ limit, in decreasing order, the three $j=N/2-1, N/2-2, N/2$ subspaces of the $\Jvec^2$ operator carry most of the information about $\varepsilon$ when $\varepsilon\ll1$, and $N\varepsilon^2\ll1$. In particular, from Eq.~(\ref{eq:pN/2}) and Eq.~(\ref{eq:pN/2-1}) we see that most detection events will be contributed by the $j=N/2$ subspace, followed by the $j=N/2-1$ subspace. It is however these rare detection events from the $j=N/2-1$ subspace that contain most information about $\varepsilon$, see Eq.~(\ref{eq:fN/2-1}); the abundant detection events from the $j=N/2$ subspace and the even rarer detection events from the $j=N/2-2$ subspace contribute only information with relative weight $N\varepsilon^2\ll1$. The identification of a few specific angular momentum `modes' that carry most information about the parameter in this regime is a reminiscent of the application of the spatial-mode demultiplexing technique for the superresolution imaging problem\cite{Tsang2016, Len2020}: at the close separation regime (corresponding to $\varepsilon\ll1$), most photons will be detected by the fundamental spatial mode---the transfer function of the imaging system (corresponding to $j=N/2$), followed by detection by the first `excited' mode, which is the derivative of the transfer function with respect to the spatial dimension (corresponding to $j=N/2-1$). Similarly there, it is these rare detection events in the first `excited' mode that carries most information and is mainly responsible for the superresolution capability of the spatial-mode demultiplexing technique. Note that, when $N$ gets ever larger such that $N\varepsilon^2>1$ and the asymptotic series approximation Eq.~(\ref{eq:pj2nd}) is no longer accurate, we expect now the information carried by the lower angular momentum modes will become increasingly important and needs to be taken into account, such that the $\Jvec^2$ measurement still provides the asymptotic optimal precision as given by Eq.~(\ref{eq:J2errpro}) and Eq.~(\ref{eq:J2errprolargeN}). All these can be summarized in Fig.~\ref{fig:MPEfQS-fig3}, where we present the FIs associated the the three largest angular momentum modes, both for the exact values and the approximate ones as given by Eqs.~(\ref{eq:fN/2}-\ref{eq:fN/2-2}), as well as their joint contribution.

\section{Realization of the $\Jvec^2$ and $L_\varphi$ measurement}\label{Sec4}
Let us now discuss the implementation of the $\Jvec^2$ and $L_\varphi$ measurement. While measuring the local operator $L_\varphi$ should be straightforward, measuring the global operator $\Jvec^2$ is more challenging. For the simplest scenario with $N=2$, exact implementation of such a scheme is known for bosonic system: measurement of the $\Jvec^2$ basis is realized using the Hong-Ou-Mandel (HOM) interference effect \cite{Hong1987, Dai2014}, whereas the $L_\varphi$ measurement is performed in its local basis, after the interference. In fact, a recent experiment demonstrating superresolution estimation of the centroid and separation between two mutually incoherent sources in the sub-Rayleigh limit  \cite{Parniak2018}---a task which can be well approximated as the estimation of $\varepsilon$ and $\varphi$ in our model \cite{Chrostowski2017}---utilizes essentially the same idea. 

For the sake of clarity as well as introducing the notation, below, we first review the known results of how the setup of HOM interferometer followed by projective measurements at the end works for the $N=2$ case. Note that here, our study include both bosons and fermions. Then, new in this work we show how this can be generalized to the $N=3$ case, before moving on to briefly discussing the more complicated scenarios with larger $N$, where the $\Jvec^2$ measurement is approached from a quantum circuit perspective, with ancilla qubits involved. Lastly, we look closer into the case of nearly-pure qubit estimation, where in conjunction with the discussion in Sec.~\ref{sec:nearly-pure}, non-exact but \emph{approximate} implementations that focus on distinguishing between just the $j=N/2$ and $j=N/2-1$ subspaces will be considered. In what follows, for convenience of the notation, we will use $\{\th, \tv\}$ to specify a basis for the qubit system, such that, e.g., $\sigma_z=\proj{\th}-\proj{\tv}$.

\subsection{$N=2$}
\begin{figure*}[t!]
	\centering
	\includegraphics[width=0.95\textwidth]{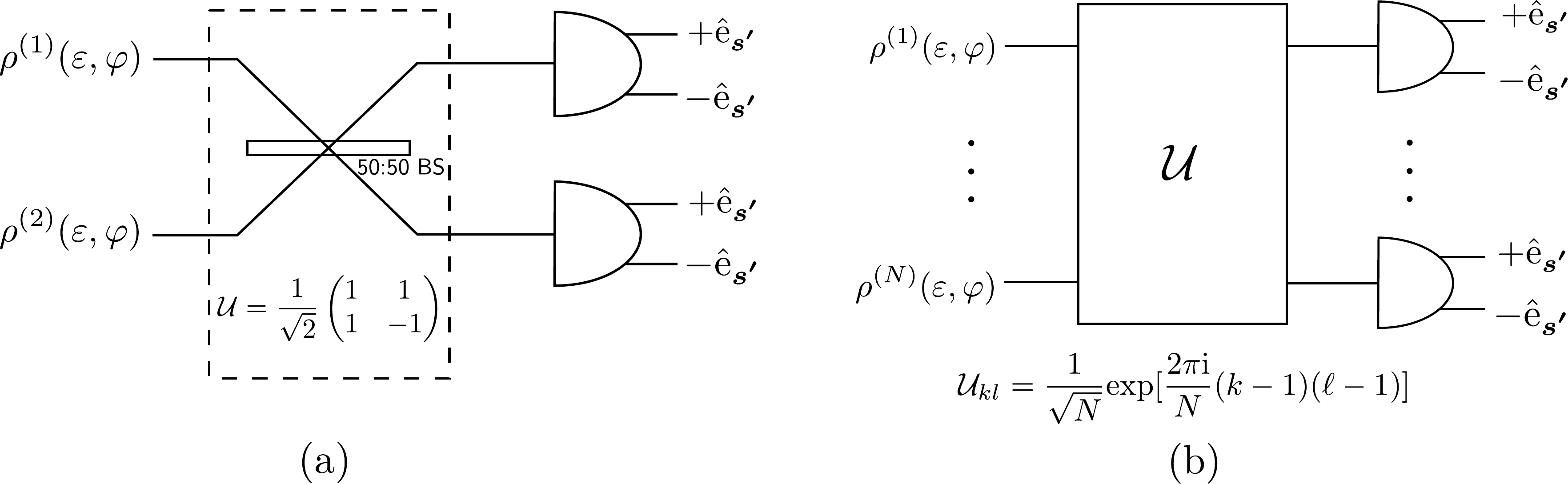}
	\caption{\textbf{(a) Implementation of $\Jvec^2$ and $L_\varphi$ measurement for $N=2$ qubits.} The two qubits are set to enter a Hong-Ou-Mandel (HOM) interferometer which consists of a 50:50 beam splitter (BS), whose action is equivalent to applying a Hadamard transformation $\mathcal{U}$ to the creation operators, see Eq.~(\ref{eq:HOMn=2}). The HOM interference signatures uniquely determine the projection onto the eigenspaces of $\Jvec^2$: bunching to the same output port is equal to $j=1(0)$ projection for bosons (fermions), while anti-bunching at different output ports is equal to $j=0(1)$ projection for bosons (fermions). At the end of each output ports, a projection onto the operators $\{\frac{\id+\hat{\upe}_{\boldsymbol{s}'}\cdot\boldsymbol{\sigma}}{2},\frac{\id-\hat{\upe}_{\boldsymbol{s}'}\cdot\boldsymbol{\sigma}}{2}\}$ then completes the measurement of $L_\varphi$. \textbf{(b) Bell multiport}: The HOM in (a) can be seen as a special case of a Bell multiport with $N=2$, where the Hadamard transformation is now generalized to the discrete Fourier transform. For $N=3$, the measurement statistics (including coincidences) at (between) the various output ports allow us to implement the $\Jvec^2$ and $L_\varphi$ measurement exactly; see Eqs.~(\ref{eq:N=3sigboson}) and (\ref{eq:N=3sigferm}). For $N\geq4$, however, Bell multiport cannot implement $\Jvec^2$ perfectly, as not all interference signatures for different $j$ values are now distinct. Nevertheless, we suggest that the interference signatures for $j=N/2$ and $j=N/2-1$ will be distinct, which we can then make use of in the limit of nearly-pure qubit estimation, as most detection events in this case are contained in this two subspaces; see Sec.~\ref{sec:nearly-pure}.}
	\label{fig:MPEfQS-fig4}
\end{figure*}

A HOM interferometer, schematically depicted in Fig.~\ref{fig:MPEfQS-fig4}(a), consists of a 50:50 beam splitter (BS), with one qubit entering simultaneously from each of the input ports. The action of the BS can be summarized as a Hadamard transformation $\mathcal{U}$ linking the input and output creation operators, i.e.,
\begin{eqnarray} \label{eq:HOMn=2}
	\left( \begin{array}{cc}
		\bd{\alpha}{1} \\
		\bd{\alpha}{2}  \end{array} \right)&=\mathcal{U}\left( \begin{array}{cc}
		\ad{\alpha}{1} \\
		\ad{\alpha}{2}  \end{array} \right)=\frac{1}{\sqrt{2}}\left( \begin{array}{cc}
		1 & 1 \\
		1 & -1  \end{array} \right)	\left( \begin{array}{cc}
		\ad{\alpha}{1} \\
		\ad{\alpha}{2}  \end{array} \right),
\end{eqnarray}
where the operator $\ad{\alpha}{i} (\bd{\alpha}{i})$ specifies the creation of a qubit with the state $\alpha=\{\th,\tv\}$ at the input (output) port $i=\{1,2\}$. As usual, we have $\big[\ad{\alpha}{i},\ad{\beta}{j}\big]=\big[\bd{\alpha}{i},\bd{\beta}{j}\big]=0$ for bosons, and $\big\{\ad{\alpha}{i},\ad{\beta}{j}\big\}=\big\{\bd{\alpha}{i},\bd{\beta}{j}\big\}=0$ for fermions, which denies the possibility of having two fermions with same degrees of freedom at the same output. In this notation, we thus have $\rhoN(\varepsilon,\varphi)=\bigotimes_{i=1}^N \rho^{(i)}(\varepsilon,\varphi)$, with
\begin{equation} \label{eq:rhoi}
\fl
\rho^{(i)}(\varepsilon,\varphi)=\upe^{-\upi \sigma_y\varphi/2}\Big(\frac{1+s(\varepsilon)}{2}\ad{\th}{i}\proj{\vac}\aa{\th}{i}+\frac{1-s(\varepsilon)}{2}\ad{\tv}{i}\proj{\vac}\aa{\tv}{i}\Big)\upe^{\upi \sigma_y\varphi/2},
\end{equation}
where $\ket{\vac}$ is the vacuum state, $\aa{\alpha}{i}=(\ad{\alpha}{i})^\dagger$. One can then show that (see \ref{app:BellmultiportN=2} for explicit demonstration with the bosonic case) we have the following unique signatures at the output ports:
\begin{eqnarray}\label{eq:N=2sigboson}
j=1&:\quad  \textrm{all two qubits exit at the same ports;}\nonumber\\
j=0&:\quad  \textrm{all two qubits exit at distinct ports,\quad \quad [bosons]}
\end{eqnarray}
or
\begin{eqnarray}\label{eq:N=2sigferm}
	j=1&:\quad  \textrm{all two qubits exit at distinct ports;}\nonumber\\
	j=0&:\quad  \textrm{all two qubits exit at the same ports.\quad [fermions]}
\end{eqnarray}
Note that these signatures are independent of the parameter $\varphi$, and hence information about it is preserved as it should. To then extract information about and estimate $\varphi$, we simply measure the projectors $\{\frac{\id+\hat{\upe}_{\boldsymbol{s}'}\cdot\boldsymbol{\sigma}}{2},\frac{\id-\hat{\upe}_{\boldsymbol{s}'}\cdot\boldsymbol{\sigma}}{2}\}$ at the end of each of the output ports, which of course corresponds to measuring the eigenspaces of $L_\varphi$. Note that, number-resolving detection technique is needed here to capture all the signatures.

\subsection{$N=3$}
A natural question arises: Can we still implement the $\Jvec^2$ and $L_\varphi$ measurement for the case of $N\geq3$, using similar idea of observing different measurement statistics, including coincidence counts, at the various output ports? Here, we attempt to answer this question by considering a generalization of HOM interferometer to the \emph{Bell multiport} setup \cite{Zukowski1997,Lim2005}, as schematically depicted in Fig.~\ref{fig:MPEfQS-fig4}(b), and can generally be realized with combinations of beam splitters, phase shifters, and mirrors \cite{Reck1994}. In this case, one qubit simultaneously enters from the each of the $N$ input of the Bell multiport, whose action, in generalization to Eq.~(\ref{eq:HOMn=2}), is the discrete Fourier transform, i.e., 
\begin{eqnarray}\label{eq:dFTN}
	\bd{\alpha}{k}=\sum_{\ell=1}^N\mathcal{U}_{k\ell}\ad{\alpha}{\ell}, \quad \quad \mathcal{U}_{k\ell}=\frac{1}{\sqrt{N}}\upe^{\frac{2\pi\upi}{N}(k-1)(\ell-1)}.
\end{eqnarray}
We work out explicitly the case for $N=3$ here, where the Bell multiport is also known as the Bell tritter \cite{Zukowski1997, Bouchard2021} and has been realized experimentally for example in photonic systems \cite{Spagnolo2013,Schaeff2015}. Leaving the details to \ref{app:BellmultiportN=3}, we find that the projection onto the $j$-eigenspace of $\Jvec^2$ has the following unique signatures at the output ports:
\begin{eqnarray}\label{eq:N=3sigboson}
	j=\frac{3}{2}&:\quad  \textrm{all three qubits exit at the same ports \emph{or}  distinct ports;}\nonumber\\
	j=\frac{1}{2}&:\quad  \textrm{two qubits exit at the same ports,}  \nonumber\\
		&\hspace{0.75cm} \textrm{another at a  different port, \quad\quad \quad [bosons]}
\end{eqnarray}
or
\begin{eqnarray}\label{eq:N=3sigferm}
	j=\frac{3}{2}&:\quad  \textrm{all three qubits exit at distinct ports;}\nonumber\\
	j=\frac{1}{2}&:\quad  \textrm{two qubits exit at the same ports,}  \nonumber\\
	&\hspace{0.75cm} \textrm{another at a  different port. \quad\;\;\quad [fermions]}
\end{eqnarray}
Thus, upon measuring the projectors $\{\frac{\id+\hat{\upe}_{\boldsymbol{s}'}\cdot\boldsymbol{\sigma}}{2},\frac{\id-\hat{\upe}_{\boldsymbol{s}'}\cdot\boldsymbol{\sigma}}{2}\}$ at the outputs, all the outcome statistics indeed allow us to implement exact $\Jvec^2$ and $L_\varphi$ measurements with a Bell tritter for $N=3$.

\subsection{$N\geq4$}
For $N\geq4$, exact realization of the measurement of $\Jvec^2$ basis using interference signatures of the ``paths" in a passive, linear setup such as a HOM or Bell tritter, to the best knowledge of the author, is not available. For example, with the Bell multiport with $N=4$ in Eq.~(\ref{eq:dFTN}), we do find that between the $j=2$ and $j=1$ eigenspaces the interference signatures are unique, but then there are no more room for another distinct signature for the $j=0$ eigenspace. In fact, as computing the output probabilities for an arbitrarily \emph{given} transformation $\mathcal{U}$ on the creation operators set $\{\ad{\alpha}{\ell}\}$ with \emph{fixed} $\alpha$ is already hard\cite{Valiant1979,Aaronson2011}---a central identity in boson sampling \cite{Aaronson2013,Brod2019}, finding a transformation $\mathcal{U}$ that has output signatures that are different for each $j$-eigenspace, which now includes also inputs of different $\alpha$, is even harder as $N$ gets larger.

Let us hence briefly switch to discuss the problem from a quantum computation perspective, where instead of just a passive setup, we allow the usage of ancillary qubits and control gates to realize the measurement of the common eigenbasis of $\Jvec^2$ and $L_\varphi$. Indeed then, upon identifying the $\Jvec^2$ (and $L_\varphi$ common) eigenbasis as the Schur basis\cite{Bacon2006}---the union of bases for both irreducible representations of the globally-symmetric unitary group as well as the permutation group---the task of projection onto this common eigenbasis can be mapped onto the outcomes of a general Schur transformation circuit, for which efficient constructions have been developed \cite{Bacon2006,Bacon2007,Kirby2017,Kirby2018}. In particular, the whole circuit can be implemented with $\Or(N^3)$ two-level gates (unitaries that act non-trivially on two dimensions), which can then be approximately executed with the universal and fault-tolerant set of Clifford and T gates \cite{Nielsen2010}. Then, should each of these two-level gates can be approximately executed with up to at most $\Or(\gamma/N^3)$ error, the full Schur transformation circuit can be realized with an error $\gamma$ using overall $\Or(N^4\log(N/\gamma))$ universal gates, as well as $\Or(\log N)$ ancilla \cite{Kirby2017,Kirby2018}.

\subsection{Nearly-pure qubit estimation: Approximate realization}
Consider now the situation in Sec.~\ref{sec:nearly-pure}, i.e., nearly-pure qubit estimation. In particular, we stick with the parameterization of $s(\varepsilon)=1-\varepsilon^2/8, \varepsilon\ll 1$.

As we have discussed in Sec.~\ref{sec:nearly-pure}, in the limit of $\varepsilon\ll1$ and $N\varepsilon^2\ll1$, the rare $j=N/2-1$ detection events are the ones that contain most information about $\varepsilon$. We can see this from another angle, by writing Eq.~(\ref{eq:rhoN}) differently, namely, up to terms of $\Or(N\varepsilon^2)$,
\begin{eqnarray}\label{eq:rhoN2}
\rhoN(\varepsilon,\varphi)&=\upe^{-\upi J_y\varphi}\Big(\frac{1+s(\varepsilon)}{2}\proj{\th}+\frac{1-s(\varepsilon)}{2}\proj{\tv}\Big)^{\otimes N}\upe^{\upi J_y\varphi}\nonumber\\
&\approx\Big(1-\frac{N\varepsilon^2}{16}\Big)\upe^{-\upi J_y\varphi}\proj{\th\cdots\th}\upe^{\upi J_y\varphi}\nonumber\\
&\hphantom{=}+\frac{\varepsilon^2}{16}\upe^{-\upi J_y\varphi}\Big(\proj{\th\cdots\th\tv}+\textsf{permutations}\Big)\upe^{\upi J_y\varphi}\nonumber\\
&\equiv w_0(\varepsilon)\tau_0(\varphi)+w_1(\varepsilon)\tau_1(\varphi),
\end{eqnarray}
where $w_0(\varepsilon)=1-\frac{N\varepsilon^2}{16}, w_1(\varepsilon)=\frac{N\varepsilon^2}{16}$, $\tau_0(\varphi)=\upe^{-\upi J_y\varphi}\proj{\th\cdots\th}\upe^{\upi J_y\varphi}$, and $\tau_1(\varphi)=\upe^{-\upi J_y\varphi}\sum_{i=1}^N\tau_{1,i}\upe^{\upi J_y\varphi}/N$ with $\tau_{1,i}=\proj{\tv}^{(i)}\bigotimes_{j(\neq i)}^N\proj{\th}^{(j)}$.

\begin{figure*}[t!]
	\centering
	\includegraphics[width=0.81\textwidth]{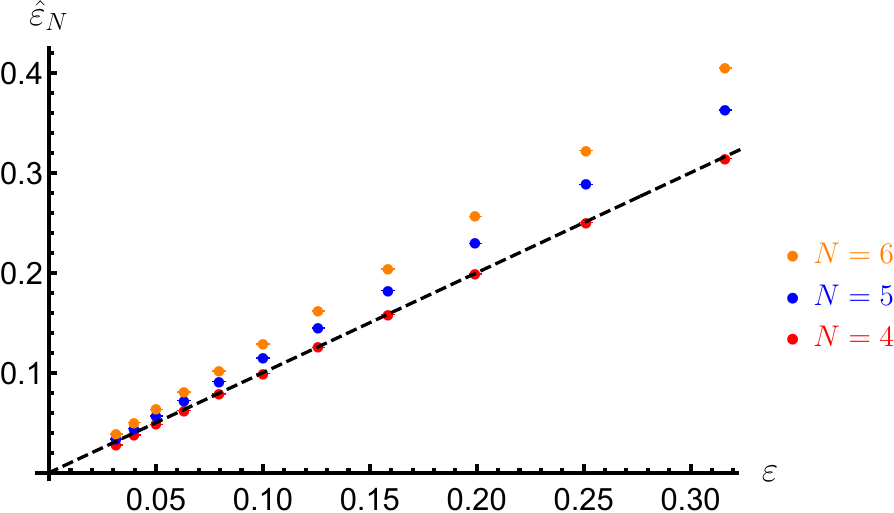}
	\caption{\textbf{Approximate estimation} of $\varepsilon$ for different $\varepsilon$ values for $N=4,5$ and 6, with the parametrization $s(\varepsilon)=1-\varepsilon^2/8$ in Eq.~(\ref{eq:rhoN}). In this simulation, each estimate is built up from $\nu=10^4$ repetitions, and we repeat the whole experiments $\mu=10^4$ times, for which their averages are plotted as the data points in the figure, and the error bars (barely visible and smaller than the markers) are constructed as the respective sample variance of these $\mu$ estimates. }
	\label{fig:MPEfQS-fig5}
\end{figure*}

Ignoring differences beyond $\Or(N\varepsilon^2)$ with the exact $\rhoN(\varepsilon,\varphi)$, we may then consider the state as given by the last line of Eq.~(\ref{eq:rhoN2}) instead, and suppose we send this state into the $N$-Bell multiport. Then, we verified explicitly up to $N=12$ that the interference signatures for all the $\tau_{1,i}$ are the same, and we denote this set of signatures by $\sig{1}$. Furthermore, our numerical results show that $\sig{1}$ and the signature set for $\tau_0(\varphi)$, $\sig{0}$, are not exactly distinct, but they have an overlap, and the probability of having an interference signature from $\tau_{1,i}$ inside this overlapping subset is given by $1/N$. The overall probability of observing $\sig{0}$ is then
\begin{eqnarray}\label{eq:prsig0}
\pr{\sig{0}}=w_0(\varepsilon)+w_1(\varepsilon)\frac{1}{N}=1-\frac{(N-1)\varepsilon^2}{16},
\end{eqnarray}
while the rest of signatures have probability
\begin{eqnarray}\label{eq:prsig1}
\pr{\sig{1}\setminus\sig{0}}=w_1(\varepsilon)(1-\frac{1}{N})=\frac{(N-1)\varepsilon^2}{16}.
\end{eqnarray}
Over $\nu$ repetitions then, should we obtain the relative frequencies $\nu_0$ and $\nu_1=1-\nu_0$ for observing signatures in $\sig{0}$ and $\sig{1}\setminus\sig{0}$ respectively, we may estimate $\varepsilon$ by linear inversion estimator as defined by
\begin{eqnarray}\label{eq:approxLinEst}
\hat{\varepsilon}(D)=4\sqrt{\frac{\nu_1}{N-1}}=\sqrt{\frac{8(1+\nu_1-\nu_0)}{N-1}},
\end{eqnarray}
which in this case is always physical, and is unbiased \emph{for the approximate state} at last line of Eq.~(\ref{eq:rhoN2}). As a reminder to the reader, by obtaining $\nu_0$ and $\nu_1$ and then estimating $\varepsilon$, we have not disturbed any information about $\varphi$ which we shall extract from $L_\varphi$ measurement, as the interference signatures are invariant under an overall and equal transformation on all the qubits. In Fig.~\ref{fig:MPEfQS-fig5}, we present a numerical simulation for the approximate estimation of different true values of $\varepsilon$ with the state $\rhoN(\thetavec)$ as given by Eq.~(\ref{eq:rhoN}) (not the approximate state) with $N=4,5,$ and 6, via the linear estimator (\ref{eq:approxLinEst}). As $N$ increases, we see that the biases in the estimator grow, which is unavoidable in this case, as now the number of angular momentum modes increases, and their contributions to both the set $\sig{0}$ and $\sig{0}\setminus\sig{1}$ increases, making the linear estimator less accurate. Evidently, this effect increases as $\varepsilon$ increases as well.

It turns out that the probability of getting the signature sets $\sig{0}$ and $\sig{1}\setminus\sig{0}$ are respectively equal to  $p_{N/2}$ and $p_{N/2-1}$, see Eqs.~(\ref{eq:prsig0}) and (\ref{eq:prsig1}) versus Eqs.~(\ref{eq:pN/2}) and (\ref{eq:pN/2-1}). Is this correspondence accidental? Evidently, $\ket{\th\cdots\th}=\ket{j=\frac{N}{2},m_z=\frac{N}{2}}$, and so $\tau_0(\varphi)=\cP_{\frac{N}{2}}\tau_0(\varphi)\cP_{\frac{N}{2}}$, such that  $\tr\{\cP_{\frac{N}{2}}\tau_0(\varphi)\}=1$. Meanwhile, we have $\tau_1(\varphi)=(\cP_{\frac{N}{2}}+\cP_{\frac{N}{2}-1})\tau_1(\varphi)(\cP_{\frac{N}{2}}+\cP_{\frac{N}{2}-1})$, and one can show easily that $\tr\{\cP_{\frac{N}{2}}\tau_1(\varphi)\}=\frac{1}{N}$. Hence, just like $\pr{\sig{0}}$ and $\pr{\sig{1}\setminus\sig{0}}$, we have $p_{N/2}\approx w_0(\varepsilon)+w_1(\varepsilon)(1/N)=1-(N-1)\varepsilon^2/16$ and $p_{N/2-1}\approx w_1(\varepsilon)(1-1/N)=(N-1)\varepsilon^2/16$, indeed. It is, therefore, inviting for us to associate the $1/N$ overlap between $\sig{1}$ and $\sig{0}$ as originated from the $1/N$ fraction of $\tau_1(\varphi)$ in the $\cP_{\frac{N}{2}}$ subspace, and therefore identify $\sig{0}$ as the signature set for $j=N/2$, while $\sig{1}\setminus\sig{0}$ as the signature set for $j=N/2-1$. 

We take note that $\sig{1}$ has been so far not obtained from any arbitrary states in the $j=N/2-1$ subspace, but only those with $m_z=N/2-1$ via the states $\tau_{1,i}=\proj{\tv}^{(i)}\bigotimes_{j(\neq i)}^N\proj{\th}^{(j)}$, $i=1,\cdots,N$. Similarly, $\sig{0}$ is obtained from considering just $\tau_0$, which is a particular state in the $j=N/2$ subspace with the specific value of $m_z=N/2$. To further confirm our assertion that $\sig{0}$ and $\sig{1}\setminus\sig{0}$ do unambiguously distinguish between the projection onto the two $j$-subspaces respectively, we check explicitly the signatures for \emph{all} the eigenstates, $\ket{j,m_z,g(j)}$, in each $(j,g(j))$-eigenspaces for $j=N/2$ and $j=N/2-1$ up to $N=6$, and find that indeed all the kets with $j=N/2$ have signatures in $\sig{0}$, while all the kets with $j=N/2-1$ have signatures in $\sig{1}\setminus\sig{0}$. We then end this section by suggesting the following from our numerical observations: 
\begin{eqnarray}\label{eq:conjecNsignatures}
\textrm{Given $N$ spin-1/2s with one spin entering simultaneously from each of } \nonumber\\
\textrm{the input ports of an $N$-Bell multiport, the interference signatures for }\nonumber\\
\textrm{the $j=N/2$ and $j=N/2-1$ subspaces are distinguishable.}
\end{eqnarray}
Put differently, we propose that an $N$-Bell multiport allows us to distinguish between projection onto the $j=N/2$ and $j=N/2-1$ subspaces of $\Jvec^2$ from the outcome signatures. Regrettably, as the numerical analysis are basically just dry computation of probabilities and sorting of outcomes, we are not able to offer any graphical explanations or any easy-to-follow arguments for the proposed statement (\ref{eq:conjecNsignatures}), other than to state we obtained the said results, and we welcome interested readers to approach the author for a copy of the numerical code.

\section{Discussion}\label{Sec5}
Ideally, we would like to have a passive implementation of the $\Jvec^2$ measurement for any valid range of $\varepsilon$ and $N$. Unfortunately however, we only manage to do so for $N=2,3$ with Bell multiport, and at the moment be content with approximate realizations in the nearly-pure qubit regime, where projection onto the $j=N/2$ and $N/2-1$ subspaces is sufficient. To proceed further, instead of Bell multiport with the corresponding discrete Fourier transform then, one may perhaps consider more general complex Hadamard transform \cite{Tadej2006}. Of course, more generally, measuring $\Jvec^2$ and $L_\varphi$ need not be the only scheme that achieves the multi-parameter QFI CRB; there might be some other schemes as well, with different implementation method altogether. We demonstrated one such scheme here and will leave these other possibilities open for now.

Regarding the interference signatures in $\sig{0}$ and $\sig{1}$ with $N$-Bell multiport in the above section, we also observe an interesting feature for the case of bosonic qubits. With $N$ bosons and $N$ outputs in the Bell multiport, there are in total $|\textsf{S}|:=\binom{2N-1}{N}$ different signatures possible. Then, we find that apart from some specific $N$, the total number of signatures in $\sig{1}$, $|\sig{1}|$, is exactly $|\textsf{S}|$, which again confirms that there are no room for other $j\neq N/2$ values to have distinguishable signature set. Up to $N=12$ as we have verified, the few exceptional $N$ values for which $|\sig{1}|\neq|\textsf{S}|$ are found to be $N=6,10,12$, with $|\sig{1}|/|\textsf{S}|=\frac{449}{462}, \frac{90358}{92378},\frac{1321717}{1352078}$ respectively. Incidentally, $N=6,10,12$ are the first three elements in the sequence of numbers which are not integer power of a prime number, and it is a well known open problem in mathematical physics and quantum information whether there exists $N+1$ mutually unbiased bases (MUB) for Hilbert space with dimension of such numbers \cite{Durt2010, Horodecki2020}. Given that one of the important tools in studying MUB is the complex Hadamard matrices for which discrete Fourier transform is one of such \cite{Tadej2006, Durt2010}, it is then perhaps not surprising to have such a correspondence between the non-unity ratio of $|\sig{1}|/|\textsf{S}|$ and the MUB problem. However, exactly how or why they are related would require a separate analysis, and is beyond the scope of this work.

\begin{figure*}[t!]
	\centering
	\includegraphics[width=0.75\textwidth]{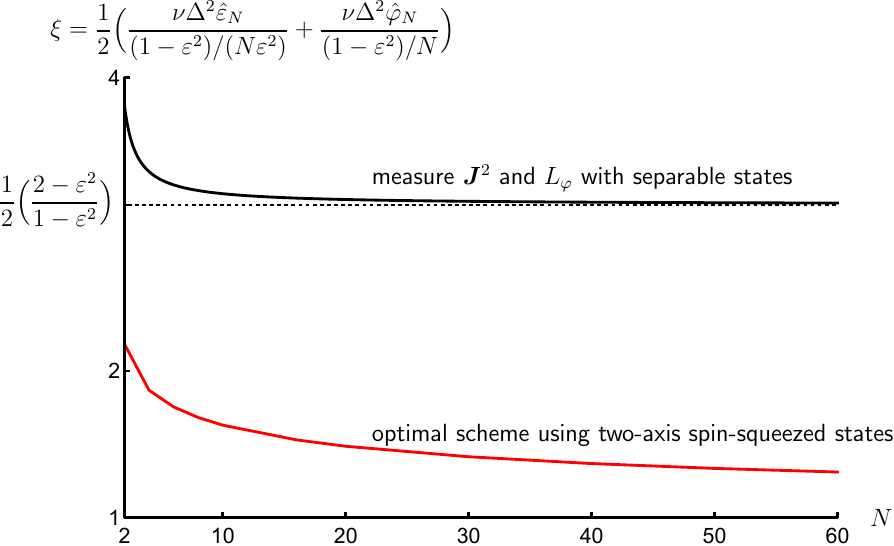}
	\caption{\textbf{$\xi$, the equal-weight sum of joint estimation MSE for the unitary phase parameter $\varphi$ and the dephasing strength parameter $\varepsilon$,} normalized by their respective optimal asymptotic precision (attainable with entangled states in general) when assuming another parameter is known and fixed (Eqs.~(\ref{eq:MSEepsas}, \ref{eq:MSEphias})), for different number of probes $N$. For the black solid curve, we consider estimation scheme using separable states with $\Jvec^2$ and $L_\varphi$ measurement introduced in Sec.~\ref{Sec3}, which has a limit of $\xi\rightarrow \frac{1}{2}\Big(\frac{2-\varepsilon^2}{1-\varepsilon^2}\Big)$ as $N\rightarrow\infty$; in the plot, we use $\varepsilon=0.9$. Except in the limit of very strong noise, $\varepsilon\rightarrow1$, we see that we cannot attain asymptotic compatible joint estimation for both parameters, i.e., $\xi\rightarrow1$. For comparison, in the red solid curve (which is also the red curve of Fig.~4 in Ref.~\cite{Ragy2016}), the corresponding $\xi$ is plotted for an optimal scheme with entangled input states such as the two-axis spin-squeezed state, which shows the trend that might possibly lead to $\xi\rightarrow1$ as $N\rightarrow\infty$.}
	\label{fig:MPEfQS-fig6}
\end{figure*}

Finally, as we have put forward in Sec.~\ref{sec:qubitmodel}, the qubit model we studied in this work can be seen as a particular case of quantum metrology or quantum channel estimation problem, in which we would like to estimate the phase parameter in an unitary encoding channel as well as the dephasing parameter in the dephasing channel, using the $N$-product input state (see Eq.~(\ref{eq:QMmodel})). More generally, one may consider entangled input states (such as one- or two axis-spin squeeze states) and general measurement strategies instead, and obtain better estimation precisions for the two parameters. Then, one may show that, when assuming one parameter is known and we estimate only the second parameter, their optimal attainable MSE in the large $N$ ensemble limit are known to be 
\begin{eqnarray}
	\Delta^2\hat{\varepsilon}_{N, \mathrm{met}}&=&\frac{1-\varepsilon^2}{N}, \label{eq:MSEepsas} \\
	\Delta^2\hat{\varphi}_{N, \mathrm{met}}&=&\frac{1-\varepsilon^2}{N\varepsilon^2}, \label{eq:MSEphias}
\end{eqnarray}	
respectively. Note that we have used the parametrization $\eta=s(\varepsilon)=\varepsilon$ here. 
	
For their joint estimation, there are no rigorous analytical results proving that the estimation of these two parameters are compatible such that these asymptotic optimal MSE can be attained simultaneously, although there are some numerical evidence suggested that this might be true \cite{Ragy2016}. Still, for comparison, let us now consider our joint estimation scheme using product input states with the collective $\Jvec^2$ and $L_\varphi$ measurement, in which 
Eq.~(\ref{eq:MSEvarphi}) gives $\Delta^2\hat{\varphi}_{N}=1/(N\varepsilon^2)$ and Eq.~(\ref{eq:J2errpro}) gives $\Delta^2\hat{\varepsilon}_{N}\sim (1-\varepsilon^2)/N$ as $N\rightarrow\infty$. This shows that we are able to attain the ultimate estimation precision $\Delta^2\hat{\varepsilon}_{N, \mathrm{met}}$ for $\varepsilon$ as $N\rightarrow\infty$, but unfortunately however there is a factor of $1/(1-\varepsilon^2)$ larger than $\Delta^2\hat{\varphi}_{N, \mathrm{met}}$---we therefore cannot attain compatible optimal estimation of these two parameters using our scheme with product input states. For illustration, in Fig.~\ref{fig:MPEfQS-fig6} we depict (black solid line) the quantity $\xi$, which is the equal-weight sum of the MSEs for our joint estimation scheme using product input states with the collective $\Jvec^2$ and $L_\varphi$ measurement, Eqs.~(\ref{eq:J2errpro}, \ref{eq:MSEvarphi}), normalized by the respective asymptotic optimal MSE Eqs.~(\ref{eq:MSEepsas}, \ref{eq:MSEphias}). Asymptotically, we have $\xi\rightarrow \frac{1}{2}\Big(\frac{1}{1-\varepsilon^2}+1\Big)=\frac{1}{2}\Big(\frac{2-\varepsilon^2}{1-\varepsilon^2}\Big)$. For comparison, we also plot the corresponding known (numerical) results of optimal joint estimation for $\varphi$ and $\varepsilon$ using the two-axis spin-squeezed states (red solid line, which is also the red curve of Fig.~4 in Ref.~\cite{Ragy2016}), which suggest that $\xi\rightarrow1$ might be attainable.

\noindent Additional remark: We realize that there is a recent work by J. O. de Almeida et al. \cite{Jessica2021} on estimation of the separation and relative intensity of two incoherent point sources, which explores similar idea of collective ``angular momentum" measurements.
\section{Conclusion}\label{Sec6}
In summary, in this work, we have studied the joint estimation of the length parameter $\varepsilon$ and the direction parameter $\varphi$ of the Bloch vector for qubit states, a model which has relevance to various physical problems, e.g., superresolution quantum imaging. We show that these two parameters can be simultaneously estimated with optimal precisions asymptotically, using the collective measurement of the angular-momentum squared operator, followed by a local projective measurement. We also discuss how such measurement scheme can be realized using Bell multiport, either exactly for $N=2,3$, or approximately when the qubits are nearly pure for other $N$ values. Finally, we propose that with a Bell multiport setup, the interference signatures in the $j=N/2$ and $j=N/2-1$ angular-momentum spaces will be distinct.


\ack {The author would like to express his gratitude to Konrad Banaszek, Chandan Datta, Marcin Jarzyna, and Jan Ko\l{}ody\'nski for their insightful comments and discussions. The author would also like to extend his appreciation to Konrad Banaszek for his encouragement and support throughout this work. The author also thanks Marcin Jarzyna for providing the data for reference in plotting Fig.~\ref{fig:MPEfQS-fig6}. This work is part of the project ``Quantum Optical Technologies" carried out within the International Research Agendas Programme of the Foundation for Polish Science co-financed by the European Union under the European Regional Development Fund.}

\appendix
\section{Precision of estimation for $\varepsilon$ with $\Jvec^2$}\label{app:errorpropJvec2}
In order to help the readers verify the expression in Eq.~(\ref{eq:J2errpro}), we list down explicitly here the various constituent terms. With $\rhoN(\varepsilon,\varphi)$ as in Eq.~(\ref{eq:rhoN}), and the notation $\expt{A}\equiv \tr\{\rhoN(\varepsilon,\varphi)A\}$ for the operator $A$, we have, by a simple rotation, or equivalently a re-definition of the $(x,y,z)$-directions, 
\begin{eqnarray}
\expt{J_x^2}&=\expt{J_y^2}=\frac{N}{4},\qquad  \expt{J_z^2}=\frac{N}{4}+\frac{N(N-1)}{4}s(\varepsilon)^2,\end{eqnarray}
such that 
\begin{eqnarray}
\expt{\Jvec^2}&=\expt{J_x^2+J_y^2+J_z^2}=\frac{3N}{4}+\frac{N(N-1)}{4}s(\varepsilon)^2, \label{eq:expJ} \\
\frac{\partial}{\partial\varepsilon}\expt{\Jvec^2}&=\frac{N(N-1)}{2}s(\varepsilon)\frac{\partial s(\varepsilon)}{\partial\varepsilon},\label{eq:tracesJ}
\end{eqnarray} 
and 
\begin{eqnarray}\label{eq:tracesJ2}
\expt{J_x^4}&=\expt{J_y^4}=\frac{N(3N-2)}{16},\nonumber \\
\expt{J_x^2J_y^2}&=\expt{J_y^2J_x^2}=\frac{N}{16}\Big(N-2(N-1)s(\varepsilon)^2\Big),\nonumber\\
\expt{J_x^2J_z^2}&=\expt{J_z^2J_x^2}=\expt{J_z^2J_y^2}=\expt{J_y^2J_z^2}=\frac{N^2}{16}\Big(1+(N-1)s(\varepsilon)^2\Big),\nonumber
\end{eqnarray}
\begin{eqnarray}\label{eq:tracesJ3}
\expt{J_z^4}=&\Big(\frac{N^4}{16}-\frac{3N^3}{8}+\frac{11N^2}{16}-\frac{3N}{8}\Big)s(\varepsilon)^4\nonumber\\
&+\Big(\frac{3N^3}{8}-\frac{7N^2}{8}+\frac{N}{2}\Big)s(\varepsilon)^2-\frac{3N^2}{16}-\frac{N}{8}, 
\end{eqnarray}
such that
\begin{eqnarray}
\fl
\expt{(\Jvec^2)^2}=\expt{J_x^4+J_y^4+J_z^4+J_x^2J_y^2+J_y^2J_x^2+J_x^2J_z^2+J_z^2J_x^2+J_y^2J_z^2+J_z^2J_y^2}\nonumber\\
\fl
=\Big(\frac{N^4}{16}-\frac{3N^3}{8}+\frac{11N^2}{16}-\frac{3N}{8}\Big)s(\varepsilon)^4+\frac{N(N-1)(5N-6)}{8}s(\varepsilon)^2-\frac{3N(5N-2)}{16}.
\end{eqnarray}
Finally then, 
\begin{eqnarray}\label{eq:tracesJ4}
\expt{(\Jvec^2)^2}-\expt{\Jvec^2}^2=\frac{N(N-1)}{8}\Big[(3-2N)s(\varepsilon)^4+2(N-3)s(\varepsilon)^2+3\Big],
\end{eqnarray}
and the MSE of estimating $\varepsilon$ in Eq.~(\ref{eq:J2errpro}) now follows immediately from Eq.~(\ref{eq:tracesJ}) and Eq.~(\ref{eq:tracesJ4}).

\section{Transformation of $\Jvec^2$ eigenstates under HOM}\label{app:BellmultiportN=2}
For $N=2$ bosonic qubits, the four $\Jvec^2$ eigenstates at the input can be chosen as
\begin{eqnarray}\label{eq:n=2input}
	\ket{j=1,m_z=1}&=\ad{\th}{1}\ad{\th}{2}\ket{\vac},\nonumber\\
	\ket{j=1,m_z=0}&=\frac{1}{\sqrt{2}}(\ad{\th}{1}\ad{\tv}{2}+\ad{\tv}{1}\ad{\th}{2})\ket{\vac},\nonumber\\
	\ket{j=1,m_z=-1}&=\ad{\tv}{1}\ad{\tv}{2}\ket{\vac},\nonumber\\
	\ket{j=0,m_z=0}&=\frac{1}{\sqrt{2}}(\ad{\th}{1}\ad{\tv}{2}-\ad{\tv}{1}\ad{\th}{2})\ket{\vac}.	
\end{eqnarray}
After traversing the HOM interferometer, these states are respectively transformed as
\begin{eqnarray}\label{eq:n=2output}
	\ket{j=1,m_z=1}&=\frac{1}{2}\left((\bd{\th}{1})^2-(\bd{\th}{2})^2\right)\ket{\vac},\nonumber\\
	\ket{j=1,m_z=0}&=\frac{1}{\sqrt{2}}(\bd{\th}{1}\bd{\tv}{1}-\bd{\th}{2}\bd{\tv}{2})\ket{\vac},\nonumber\\
	\ket{j=1,m_z=-1}&=\frac{1}{2}\left((\bd{\tv}{1})^2-(\bd{\tv}{2})^2\right)\ket{\vac},\nonumber\\
	\ket{j=0,m_z=0}&=\frac{1}{\sqrt{2}}(\bd{\tv}{1}\bd{\th}{2}-\bd{\th}{1}\bd{\tv}{2})\ket{\vac}.	
\end{eqnarray}
From Eq.~(\ref{eq:n=2output}), we thus see that the projection onto the respective $j$-eigenspace of $\Jvec^2$ can be identified with the  unique signatures as stated in Eq.~(\ref{eq:N=2sigboson}). Similar analysis can be done for fermions to obtain Eq.~(\ref{eq:N=2sigferm}).

\section{Transformation of $\Jvec^2$ eigenstates under Bell tritter}\label{app:BellmultiportN=3}
For $N=3$ bosonic qubits, a choice of the $j=\frac{3}{2}$-eigenstates for $\Jvec^2$ at the input is
\begin{eqnarray}\label{eq:n=3inputj=3/2}
\ket{j=\frac{3}{2},m_z=\frac{3}{2}}&= \ad{\th}{1}\ad{\th}{2}\ad{\th}{3}\ket{\vac},\nonumber\\
\ket{j=\frac{3}{2},m_z=\frac{1}{2}}&= \frac{1}{\sqrt{3}}(\ad{\th}{1}\ad{\th}{2}\ad{\tv}{3}+\ad{\th}{1}\ad{\tv}{2}\ad{\th}{3}+\ad{\tv}{1}\ad{\th}{2}\ad{\th}{3})\ket{\vac},\nonumber\\
\ket{j=\frac{3}{2},m_z=\frac{-1}{2}}&= \ket{j=\frac{3}{2},m_z=\frac{1}{2}}\Bigg|_{\th\leftrightarrow\tv},\nonumber\\
\ket{j=\frac{3}{2},m_z=\frac{-3}{2}}&=\ket{j=\frac{3}{2},m_z=\frac{3}{2}}\Bigg|_{\th\leftrightarrow\tv}.	
\end{eqnarray}
We focus on the $m_z=\frac{3}{2}$ and $m_z=\frac{1}{2}$ kets, as the other two are equivalent to them up to interchanging $\th\leftrightarrow\tv$, and thus will have the same interference signatures by a Bell tritter. Straightforward algebra then leads us to 
\begin{eqnarray}\label{eq:n=3outputj=3/2}
\fl
\ket{j=\frac{3}{2},m_z=\frac{3}{2}}=&\frac{1}{3\sqrt{3}}\Big((\bd{\th}{1})^3+(\bd{\th}{2})^3+(\bd{\th}{3})^3 -3\bd{\th}{1}\bd{\th}{2}\bd{\th}{3}\Big) \ket{\vac},\nonumber\\ \fl
\ket{j=\frac{3}{2},m_z=\frac{1}{2}}=& \frac{1}{3}\Big((\bd{\th}{1})^2\bd{\tv}{1}+(\bd{\th}{2})^2\bd{\tv}{2}+(\bd{\th}{3})^2\bd{\tv}{3}\nonumber\\
&-\bd{\th}{1}\bd{\th}{2}\bd{\tv}{3}-\bd{\th}{1}\bd{\tv}{2}\bd{\th}{3}-\bd{\tv}{1}\bd{\th}{2}\bd{\th}{3}\Big)\ket{\vac}.
\end{eqnarray}
We thus see that the signatures for $j=\frac{3}{2}$ after traversing the Bell tritter are either all qubits exit at the same output ports, or all exit at different output ports.

For $j=\frac{1}{2}$, there are $\mu_{j}=2$ subspaces. The first one for example has the eigenkets
\begin{eqnarray}\label{eq:n=3inputj=1/2-1}
\fl 
\ket{j=\frac{1}{2},m_z=\frac{1}{2}, g(j)=1}&= \Big(\frac{-1}{\sqrt{6}}(\ad{\th}{1}\ad{\tv}{2}+\ad{\tv}{1}\ad{\th}{2})\ad{\th}{3}+\sqrt{\frac{2}{3}}\ad{\th}{1}\ad{\th}{2}\ad{\tv}{3}\Big)\ket{\vac}, \nonumber\\
\fl \ket{j=\frac{1}{2},m_z=\frac{-1}{2}, g(j)=1}&= -\ket{j=\frac{1}{2},m_z=\frac{1}{2}, g(j)=1}\Bigg|_{\th\leftrightarrow\tv}.
\end{eqnarray}
The $m_z=\frac{1}{2}$ state, upon introducing $u=\upe^{2\pi\upi/3}$, is also
\begin{eqnarray}\label{eq:n3outputj=1/2-1}
\fl &\frac{1}{\sqrt{18}}\Big(u(\bd{\th}{1})^2\bd{\tv}{2}+u^*(\bd{\th}{1})^2\bd{\tv}{3}+ u^*(\bd{\th}{2})^2\bd{\tv}{1}+u(\bd{\th}{2})^2\bd{\tv}{3}+u(\bd{\th}{3})^2\bd{\tv}{1}\nonumber\\
\fl &+ u^*(\bd{\th}{3})^2\bd{\tv}{2}-u\bd{\th}{1}\bd{\tv}{1}\bd{\th}{2}-u^*\bd{\th}{1}\bd{\tv}{1}\bd{\th}{3}-u^*\bd{\th}{2}\bd{\tv}{2}\bd{\th}{1}-u\bd{\th}{2}\bd{\tv}{2}\bd{\th}{3}\nonumber\\
\fl &-u\bd{\th}{3}\bd{\tv}{3}\bd{\th}{1}-u^*\bd{\th}{3}\bd{\tv}{3}\bd{\th}{2}\Big)\ket{\vac},
\end{eqnarray}
which has the signature of two qubits exit at the same output ports, and one at another. The same conclusion holds for the $m_z=-\frac{1}{2}$ state.

Lastly, for the eigenkets in the second $j=\frac{1}{2}$-subspace, we have 
\begin{eqnarray}\label{eq:n=3inputj=1/2-2}
\fl 
\ket{j=\frac{1}{2},m_z=\frac{1}{2}, g(j)=2}&= \Big(\frac{1}{\sqrt{2}}(\ad{\th}{1}\ad{\th}{2}-\ad{\tv}{1}\ad{\th}{2})\ad{\th}{3}\Big)\ket{\vac}, \nonumber\\
\fl \ket{j=\frac{1}{2},m_z=\frac{-1}{2}, g(j)=2}&= -\ket{j=\frac{1}{2},m_z=\frac{1}{2}, g(j)=2}\Bigg|_{\th\leftrightarrow\tv}.
\end{eqnarray}
The $m_z=\frac{1}{2}$ state is also
\begin{eqnarray}\label{eq:n3outputj=1/2-2}
\fl &\frac{\upi}{\sqrt{18}}\Big(u(\bd{\th}{1})^2\bd{\tv}{2}-u^*(\bd{\th}{1})^2\bd{\tv}{3}- u^*(\bd{\th}{2})^2\bd{\tv}{1}+u(\bd{\th}{2})^2\bd{\tv}{3}+u(\bd{\th}{3})^2\bd{\tv}{1}\nonumber\\
\fl &- u^*(\bd{\th}{3})^2\bd{\tv}{2}-u\bd{\th}{1}\bd{\tv}{1}\bd{\th}{2}+u^*\bd{\th}{1}\bd{\tv}{1}\bd{\th}{3}+u^*\bd{\th}{2}\bd{\tv}{2}\bd{\th}{1}-u\bd{\th}{2}\bd{\tv}{2}\bd{\th}{3}\nonumber\\
\fl &-u\bd{\th}{3}\bd{\tv}{3}\bd{\th}{1}+u^*\bd{\th}{3}\bd{\tv}{3}\bd{\th}{2}\Big)\ket{\vac},
\end{eqnarray}
which again has the signature of two qubits exiting at the same output ports, and one at another, and the same conclusion holds for the $m_z=-\frac{1}{2}$ state. Therefore, overall, the projections onto $j=\frac{1}{2}$ subspaces all have the same interference signature, which is perfectly distinguishable from the signatures associated with the $j=\frac{3}{2}$ subspace as summarized in Eq.~(\ref{eq:N=3sigboson}). Similar analysis can be done for fermions to obtain Eq.~(\ref{eq:N=3sigferm}).


\end{document}